\documentclass[aps,prl,twocolumn,superscriptaddress,showpacs]{revtex4}
\usepackage{placeins}
\usepackage{graphicx}
\usepackage{amsmath}
\usepackage{amssymb}
\usepackage{times}
\usepackage{color}

\begin{document}

\preprint{}

\title{Spatiotemporal complexity of a ratio-dependent predator-prey system}

\author{Weiming Wang}
\email{weimingwang2003@163.com} \affiliation{Department of
Mathematics, North University of China, Taiyuan, Shan'xi 030051,
P.R. China} \affiliation{School of Mathematics and Information
Science, Wenzhou University, Wenzhou,Zhejiang, 325035}%
\author{Quan-Xing Liu}\email{liuqx315@sina.com}
\affiliation{Department of Mathematics, North University of China,
Taiyuan, Shan'xi 030051, P.R. China}
\author{Zhen Jin}\email{jinzhn@263.net}
\affiliation{Department of Mathematics, North University of China,
Taiyuan, Shan'xi 030051, P.R. China}

\date{\today}

\begin{abstract}
In this paper, we investigate the emergence of a ratio-dependent
predator-prey system with Michaelis-Menten-type functional response
and reaction-diffusion. We derive the conditions for Hopf, Turing
and Wave bifurcation on a spatial domain. Furthermore, we present a
theoretical analysis of evolutionary processes that involves
organisms distribution and their interaction of spatially
distributed population with local diffusion. The results of
numerical simulations reveal that the typical dynamics of population
density variation is the formation of isolated groups, i.e.,
stripelike or spotted or coexistence of both. Our study shows that
the spatially extended model has not only more complex dynamic
patterns in the space, but also chaos and spiral waves. It may help
us better understand the dynamics of an aquatic community in a real
marine environment.

\end{abstract}


\maketitle

\section{Introduction}

Ecological systems are characterized by the interaction between
species and their natural environment~\cite{Martin2006}. Such
interaction may occur over a wide range of spatial and temporal
scales~\cite{ClaudiaNeuhauser,SimonA.Levin01171997}. The study of
complex population dynamics is nearly as old as population ecology.
In the 1920s, Lotka and Volterra independently developed a simple
model of interacting species that still bears their joint names.
This was a nearly linear model, but the predator-prey version
displayed neutrally stable cycles~\cite{May1976,Kendall2001}. From
then on, the dynamic relationship between predators and their prey
has long been and will continue to be one of dominant themes in both
ecology and mathematical ecology due to its universal existence and
importance~\cite{Berryman,kuang98global,Murray2004}.

Predator-prey models follow two general principles: one is that
population dynamics can be decomposed into birth and death
processes; the other is the conservation of mass principle, stating
that predators can grow only as a function of what they have
eaten~\cite{Jost1998}. With these two principles we can write the
canonical form of a predator-prey system as
\begin{eqnarray}\label{eq:1}
\left\{ \begin{array}{l}
 \dot x(t)=xg(x)-f(x,y)y-\mu_x(x) x, \\[8pt]
 \dot y(t)=\gamma f(x,y)y- \mu_y(y) y. \\
 \end{array} \right.
\end{eqnarray}
where $g(x)$ is the per capita prey growth rate in the absence of
the predator, $\mu_x$ and $\mu_y$ are natural mortalities of prey
and predator respectively, $f(x,y)$ is the functional response. And
$\gamma f(x,y)$ is the per capita production of predator due to
predation, which is often called the numerical response. The
functional response plays a main role in system~\eqref{eq:1}: the
knowledge of this function determines the dynamics of the whole
system and the transfer of the biomass in the predation because it
is proportional to the numerical response. Usually one considers
consumption to be the major death cause for the prey. In this case
$\mu_x(x)$ can be neglected and set to 0 (as long as the predator
exists)~\cite{Jost1998}.

In population dynamics, a functional response of the predator to the
prey density refers to the change in the density of prey attached
per unit time per predator as the prey density
changes~\cite{Ruan:1445}. In general, functional response can be
classified as: (i) prey dependent, when prey density alone
determines the response, i.e., $f(x,y)=p(x)$; (ii) predator
dependent, when both predator and prey populations affect the
response. Particularly, when $f(x,y)=p(\frac{x}{y})$, we call
model~\eqref{eq:1} strictly ratio-dependent; and (iii) multi-species
dependent, when species other than the focal predator and its prey
species influence the functional response~\cite{Abram2000}.
Differing from the prey-dependent predator-prey models, the
ratio-dependent predator-prey systems have two principal
predictions: (a) equilibrium abundances are positively correlated
along a gradient of enrichment and (b) the ``paradox of enrichment"
either completely disappears or enrichment is linked to stability in
a more complex way~\cite{Arditi1989,MichaelL.Rosenzweig01291971}.
The ratio-dependent predator-prey model has been studied by several
researchers recently and very rich dynamics have been
observed~\cite{Arditi1989,kuang98global,Jost1999,Murray2004,Xiao2005}.

On the other hand, pattern formation in nonlinear complex systems is
one of the central problems of the natural, social, and
technological sciences~\cite{medvinsky:311}. In particular, starting
with the pioneering work of Segel and Jackson~\cite{Segel1972},
spatial patterns and aggregated population distributions are common
in nature and in a variety of spatio-temporal models with local
ecological interactions~\cite{Martin2006,Pascual1993}. Promulgated
by the theoretical paper of Turing~\cite{Turing1952}, the field of
research on pattern formation modeled by reaction-diffusion systems,
which provides a general theoretical framework for describing
pattern formation in systems from many diverse disciplines including
(but not limited to)
biology~\cite{Klausmeier06111999,medvinsky:311,PhysRevE.74.011914,Hardenberg,
PhysRevE.66.010901,Sharon2004,Pascual},
chemistry~\cite{Ouyang1991,Vanag2000,yang:7259,PhysRevLett.88.208303,
yang:178303,yang:026211},
physics~\cite{PhysRevE.64.026219,RevModPhys.65.851,yochelis:236,PhysRevE.70.066202},
and so on, seems to be a new increasingly interesting area,
particularly during the last decade.

In Ref.~\cite{ClaudiaNeuhauser}, Neuhauser surveys some current work
on spatial mathematical models in ecology. Much of this work
consists of building spatial dimensions into existing classical
models, such as the Lotka-Volterra model that describes competition
between species. But the research on the spatial pattern of
ratio-dependent predator-prey models seems to be rare.

\section{Stability and Bifurcation analysis}

In this paper, we mainly focus on the ratio-dependent predator-prey
system with Michaelis-Menten-type(or Michaelis-Menten-Holling)
functional response:
\begin{eqnarray}\label{eq:2}
\left\{ \begin{array}{l}
 \frac{\partial N}{\partial t}=r(1-\frac{N}{K})N-\frac{\alpha N/P}{1+\alpha hN/P}P+D_1\nabla^2N\\[8pt]
 \quad\,\,\,\,=r(1-\frac{N}{K})N-\frac{\alpha N}{P+\alpha hN}P+D_1\nabla^2N, \\[8pt]
 \frac{\partial P}{\partial t}=\gamma\frac{\alpha N}{P+\alpha hN}P-\mu P+D_2\nabla^2P.
 \\[8pt]
 \forall (N,P)\in[0, \infty]^2\backslash(0,0).
 \end{array} \right.
\end{eqnarray}
where $N, P$ stand for prey and predator density, respectively.
$D_1, D_2$ are their respective diffusion coefficients,
$\nabla^2=\frac{\partial}{\partial x^2}+\frac{\partial}{\partial
y^2}$. All parameters are positive constants, $r$ stands for maximal
growth rate of the prey, $\gamma$ conversion efficiency, $\mu$
predator death rate, $K$ carrying capacity, $\alpha$ capture rate
and $h$ handling time.

Note that $\frac{\alpha N/P}{1+\alpha hN/P}$ is strictly correct
only for $P>0$. In the case of $P=0$ and $N>0$ we can define
$f(N,0):=\frac{1}{h}$ (the limit of $f(x)$ for $x\rightarrow
\infty$).

Let \begin{equation}\label{eq:3}
\begin{array}{l} \hat{N}=\frac{\alpha h
N}{\gamma K},\quad
\hat{P}=\frac{\alpha h N}{\gamma^2 K}, \quad R=\frac{rh}{\gamma},\\[8pt]
Q=\frac{h\mu}{\gamma},\quad S=\frac{\alpha h}{\gamma},\quad
\hat{t}=\frac{\gamma t}{h}.
\end{array}
\end{equation}
For simplicity we will not write the hat $(\,\hat{}\,)$ in the rest
of this paper. And in these new variables, from
\eqref{eq:2}-\eqref{eq:3}, we arrive at the following equations
containing dimensionless quantities:
\begin{eqnarray}\label{eq:4}
\left\{ \begin{array}{l}
 \frac{\partial N}{\partial t}=R(1-\frac{N}{S})N-\frac{SN}{P+SN}P+D_1\nabla^2N, \\[8pt]
 \frac{\partial P}{\partial t}=\frac{SN}{P+SN}P-QP+D_2\nabla^2P. \\
 \end{array} \right.
\end{eqnarray}
More details about the choice of dimensionless variables in the
system \eqref{eq:2} as well as possible implications can be found in
~\cite{Jost1999}.

The dimensionless model (Eq.~\ref{eq:4}) has only three parameters:
$R$, which controls the growth rate of prey; $Q$, which controls the
death rate of the predator; and $S$, which measures capturing rate.

The first step in analyzing the model is to determine the behavior
of the non-spatial model obtained by setting space derivatives equal
to zero. The non-spatial model has at most three equilibria
(stationary states), which correspond to spatially homogeneous
equilibria of the full model (Eq.~\ref{eq:4}), in the positive
quadrant: $(0, 0)$ (total extinct), $(S, 0)$ (extinct of the
predator) and a nontrivial stationary state $(n^*, p^*)$
(coexistence to prey and predator), where
\begin{equation}\label{eq:5}
\begin{array}{l}
n^*=\frac{S(R+(Q-1)S)}{R},\\[8pt]
p^*=\frac{S(1-Q)}{Q}n^*=\frac{{S}^{2}(R-S+QS)(1-Q)}{RQ}
\end{array}
\end{equation}
Easy to know that $n^*$ is positive for all $S<\frac{R}{1-Q}$, which
implies $Q<1$ and therefore ensures the positivity of
$p^*$~\cite{Jost1999}.

To perform a linear stability analysis, we linearize the dynamic
system~\eqref{eq:4} around the spatially homogenous fixed
point~\eqref{eq:5} for small space- and time-dependent fluctuations
and expand them in Fourier space
\begin{equation}\label{eq:6}
\begin{array}{l}
N(\vec{x},t)\sim n^*e^{\lambda t}e^{i\vec{k}\cdot\vec{x}},\\[8pt]
P(\vec{x},t)\sim p^*e^{\lambda t}e^{i\vec{k}\cdot\vec{x}}.
\end{array}
\end{equation}
and obtain the characteristic equation
\begin{equation}\label{eq:7}
|A-k^2D-\lambda I|=0,
\end{equation}
where
\begin{equation}\label{eq:8}
D=\left(\begin {array}{cc}D_1&0\\\noalign{\medskip}0&D_2\end {array}
\right),
\end{equation}
and $A$ is given by
\begin{equation}\label{eq:9}
A=\left(\begin
{array}{cc}\partial_Nf&\partial_Pf\\\noalign{\medskip}\partial_Ng&\partial_Pg\end
{array} \right)_{(n^*,p^*)}=\left(\begin
{array}{cc}f_N&f_P\\\noalign{\medskip}g_N&g_P\end {array} \right),
\end{equation}
where the elements are the partial derivatives of the reaction
kinetics evaluated at the stationary state $(n^*, p^*)$. Now
Eq.~\eqref{eq:7} can be solved, yielding the so called
characteristic polynomial of the original problem (Eq.~\ref{eq:4})
\begin{equation}\label{eq:10}
\lambda^2-tr_k\lambda+\Delta_k=0,
\end{equation}
where
\begin{equation}
tr_k=f_N+g_P-k^2(D_1+D_2)=tr_0-k^2(D_1+D_2),
\end{equation}
\begin{equation}\begin{array}{l}
\Delta_k=f_Ng_P-f_Pg_N-k^2(f_ND_2+g_PD_1)+k^4D_1D_2\\[8pt]
\quad\,\,\,\,=\Delta_0-k^2(f_ND_2+g_PD_1)+k^4D_1D_2,
\end{array}\end{equation}
The roots of Eq.10 yield the dispersion relation
\begin{equation}
\lambda_{1,2}(k)=\frac{1}{2}\Bigl(tr_k\pm\sqrt{tr_k^2-4\Delta_k}\,\,\Bigr).
\end{equation}

It's well known that reaction-diffusion systems have led to the
characterization of three basic types of symmetry-breaking
bifurcations responsible for the emergence of spatiotemporal
patterns. The space-independent Hopf bifurcation breaks the temporal
symmetry of a system and gives rise to oscillations that are uniform
in space and periodic in time. The (stationary) Turing bifurcation
breaks spatial symmetry, leading to the formation of patterns that
are stationary in time and oscillatory in space. The wave
(oscillatory Turing or finite-wavelength Hopf) bifurcation breaks
both spatial and temporal symmetry, generating patterns that are
oscillatory in space and time~\cite{yang:7259}.

The Hopf bifurcation occurs when
\begin{equation}
Im(\lambda(k))\neq 0, \quad Re(\lambda(k))=0\,\,\,at\,\,\,k=0.
\end{equation}
then we can get the critical value of Hopf bifurcation parameter $S$
equals
\begin{equation}
S_H=\frac{R+Q-Q^2}{1-Q^2}.
\end{equation}
At the Hopf bifurcation threshold, the temporal symmetry of the
system is broken and gives rise to uniform oscillations in space and
periodic oscillations in time with the frequency
$$\omega_H=Im(\lambda(k))=\sqrt{\Delta_0}=\sqrt{Q(Q-1)(R-S+QS)},$$
the corresponding wavelength is
\begin{equation}
\lambda_H=\frac{2\pi}{\omega_H}=\frac{2\pi}{\sqrt{Q(Q-1)(R-S+QS)}}.
\end{equation}

The Turing bifurcation occurs when
\begin{equation}
Im(\lambda(k))=0, \quad Re(\lambda(k))=0\,\,\,at\,\,\,k=k_T\neq 0.
\end{equation}
the critical value of bifurcation parameter $S$ equals
\begin{equation}
S_T=\frac{D_1D_2k_T^4+(D_2R+D_1Q(1-Q))k_T^2+RQ(1-Q)}{Q^3
-(k_T^2D_2+2)Q^2+Q+k_T^2D_2},
\end{equation}
where
$$k_T^2=\sqrt{\frac{\Delta_0}{D_1D_2}},$$
and at the Turing threshold, the spatial symmetry of the system is
broken and the patterns are stationary in time and oscillatory in
space with the wavelength
\begin{equation}
\lambda_T=\frac{2\pi}{k_T}.
\end{equation}

And the Wave bifurcation occurs when
\begin{equation}
Im(\lambda(k))\neq 0, \quad Re(\lambda(k))=0\,\,\,at\,\,\,k=k_w\neq
0.
\end{equation}
the critical value of Wave bifurcation parameter $S$ equals
\begin{equation}
S_W=\frac{k_w^2(D_1+D_2)+R+Q-Q^2}{1-Q^2},
\end{equation}
where
$$\begin{array}{l}k_w^2=\frac{Q}{2D_2^2(Q+1)}\Bigl((D_1-D_2)^2Q^4-2(D_1^2+D_2^2)Q^3
\Bigr.\\[8pt]+(D_1^2+D_2^2
\Bigl.+6D_1D_2-4D_2^2R)Q^2-4D_1D_2Q+4D_2^2R\Bigr)^{1/2}.\end{array}$$
Easy to know that, at the Wave threshold, both spatial and temporal
symmetries are broken and the patterns are oscillatory in space and
time with the wavelength
\begin{equation}
\lambda_W=\frac{2\pi}{k_w}.
\end{equation}

\begin{figure}[htp]
\includegraphics[width=6cm]{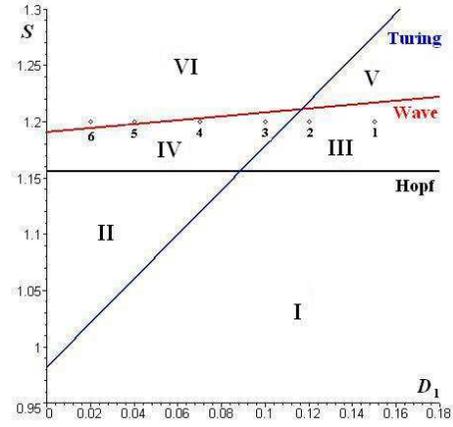}
\caption{\label{fig1}(Color online) Bifurcation diagram for the
system (2) with $R=0.5,\,Q=0.6,\,D_2=0.2$. Hopf bifurcation line:
$S_H=\frac{37}{32}$; Turing bifurcation line:
$S_T=\frac{55}{56}+\frac{55}{28}D_1$ with $k_T^2=5.48571$; Wave
bifurcation line:
$S_W=(\frac{25}{16}D_1+\frac{5}{16})k_W^2+\frac{37}{32}$ with
$k_W=0.334$. Turing-Hopf bifurcation point: (0.08864, 1.15625) and
Turing-Wave bifurcation point: (0.11655, 1.21110).}
\end{figure}

Linear stability analysis yields the bifurcation diagram with
$R=0.5,\,Q=0.6,\,D_2=0.2$ shown in Fig.~\ref{fig1}.

\begin{figure}[htp]
\includegraphics[width=6cm]{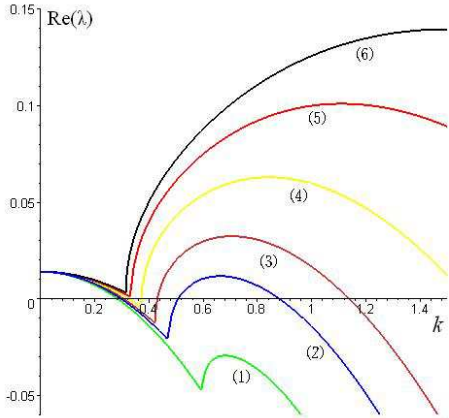}
\caption{\label{fig2}(Color online) Dispersion relations showing
unstable Hopf mode, transition of Turing and Wave modes from stable
to unstable, e.g., as $D_1$ decreased. Parameters:
$S=1.2,\,R=0.5,\,Q=0.6,\,D_2=0.2$ and (1) $D_1$=0.15; (2)
$D_1$=0.12; (3) $D_1$=0.10; (4) $D_1$=0.07; (5) $D_1$=0.04; (6)
$D_1$=0.02.}
\end{figure}

The Hopf bifurcation line, the Wave bifurcation line and the Turing
bifurcation line intersect at two codimension-2 bifurcation points,
the Turing-Hopf bifurcation point and the Turing-Wave bifurcation
point. The bifurcation lines separate the parametric space into six
distinct domains. In domain I, located below all three bifurcation
lines, the steady state is the only stable solution of the system.
Domain II is region of pure Turing instabilities, and domain III is
pure Hopf instabilities. In domain IV, both Hopf and Turing
instabilities occur, and in domain V, the Wave and Hopf modes arise.
When the parameters correspond to domain VI, which is located above
all three bifurcation lines, all three instabilities occur.
Fig.~\ref{fig2} shows the dispersion relations of unstable Hopf
mode, transition of Turing and Wave modes from stable to unstable.
It's easy to know that all three bifurcations are supercritical.

\section{Spatiotemporal pattern analysis}

\begin{figure}[htp]
\includegraphics[width=3cm]{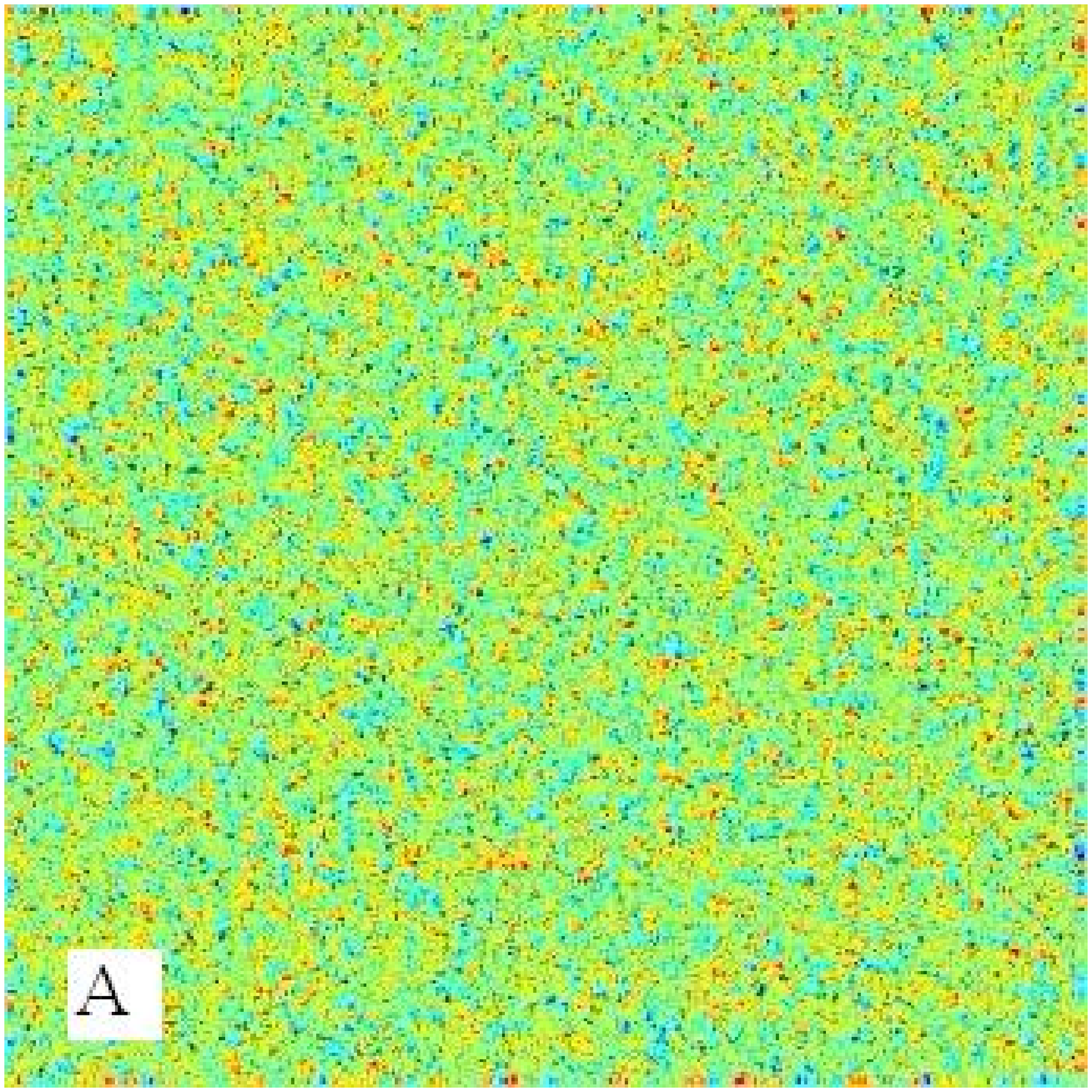}
\includegraphics[width=3cm]{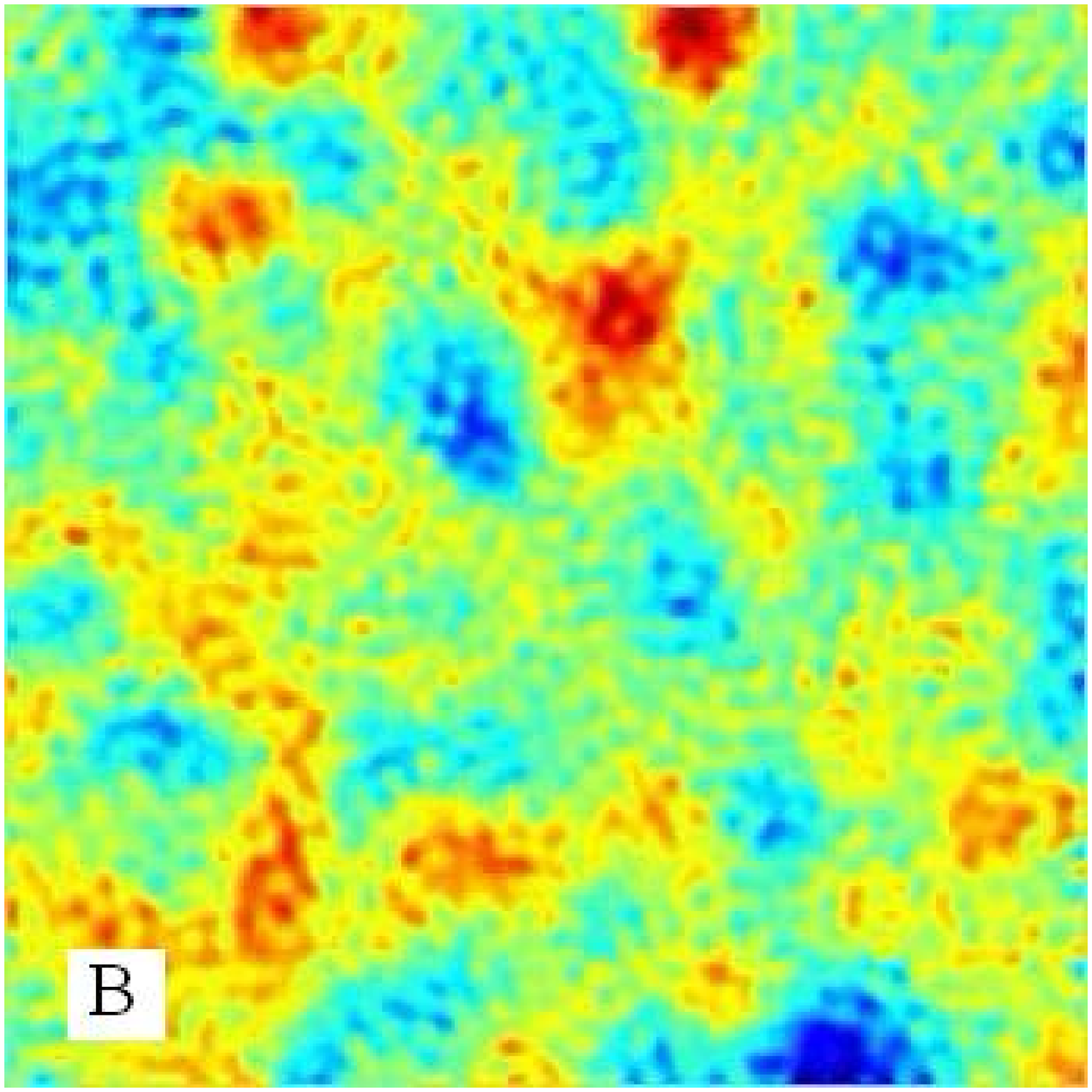}
\includegraphics[width=3cm]{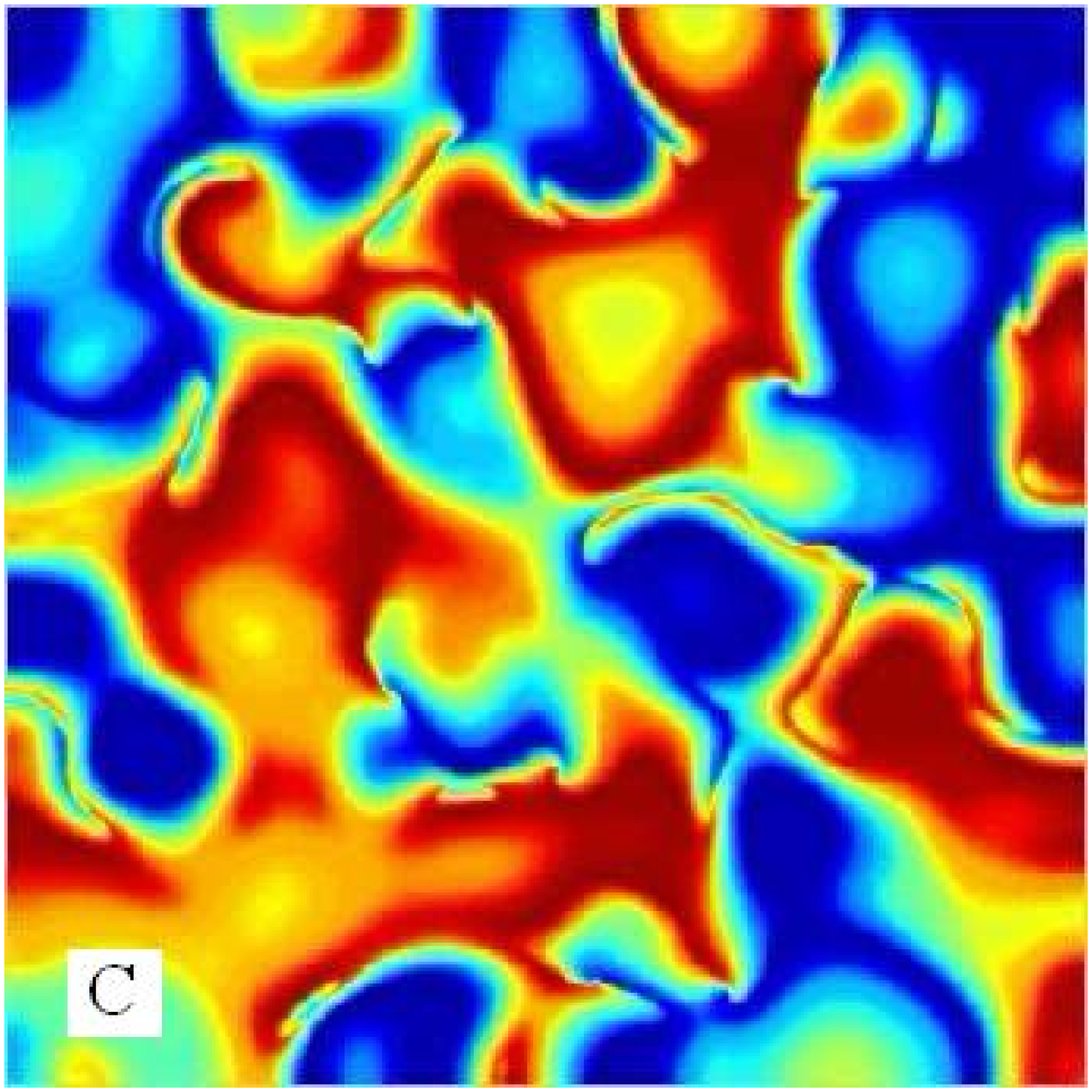}
\caption{\label{figR1}(Color online) Snapshots of contour pictures
of the time evolution of the prey at different instants with
$S=0.6<S_{T}$. (A) 0 iteration; (B) 5000 iterations; (C) 45000
iterations. [Additional movie format available from the author]}
\end{figure}

\begin{figure}[htp]
\includegraphics[width=3cm]{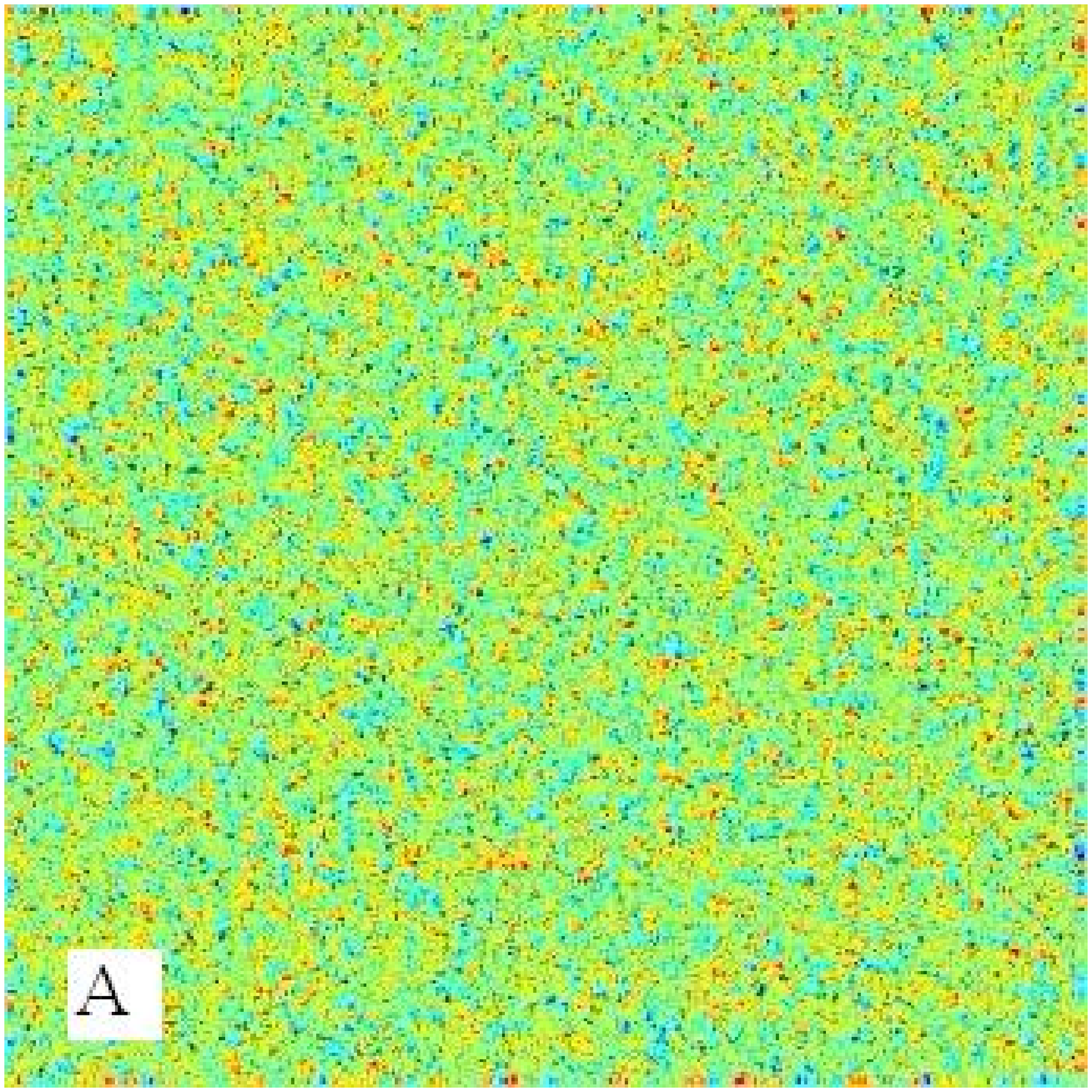}
\includegraphics[width=3cm]{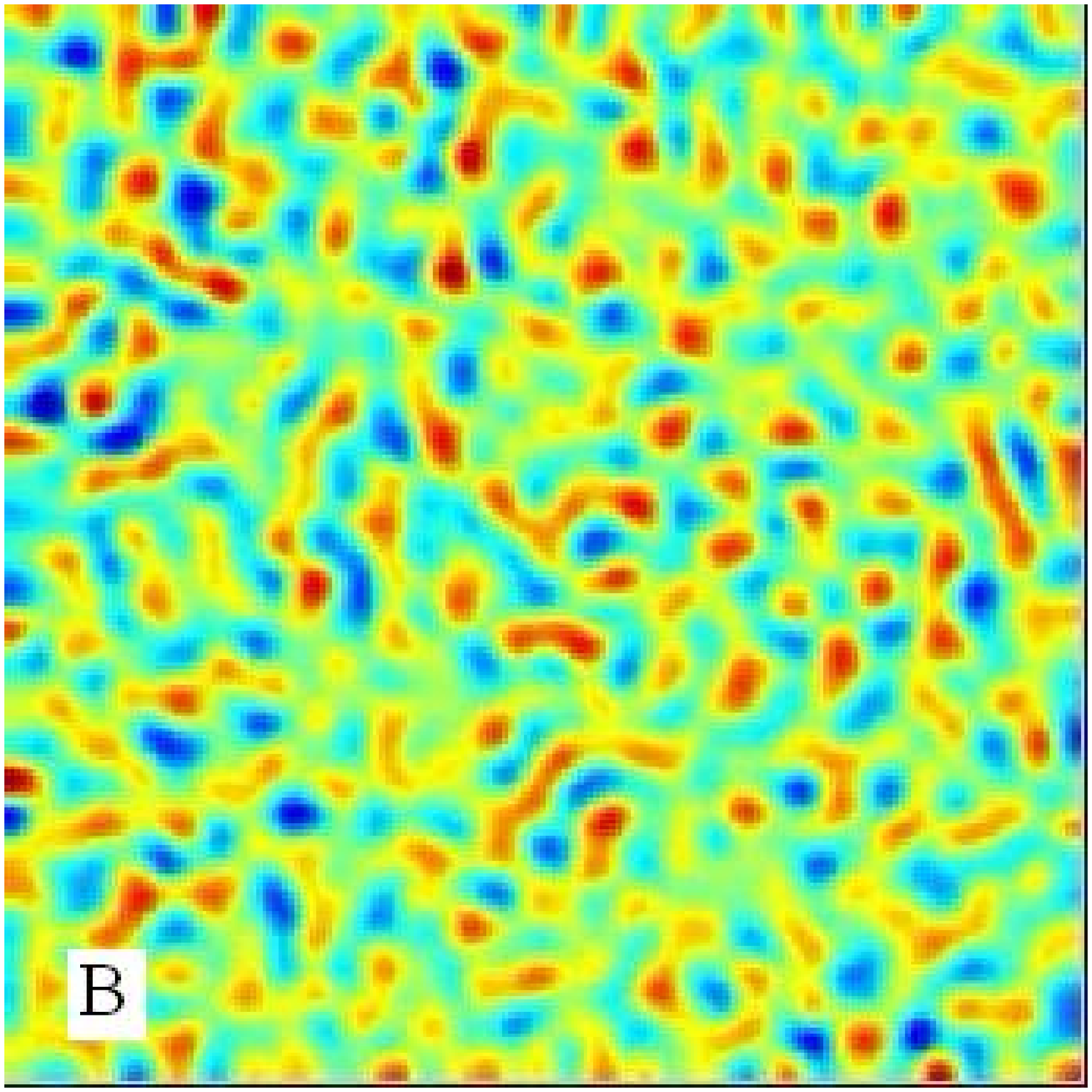}
\includegraphics[width=3cm]{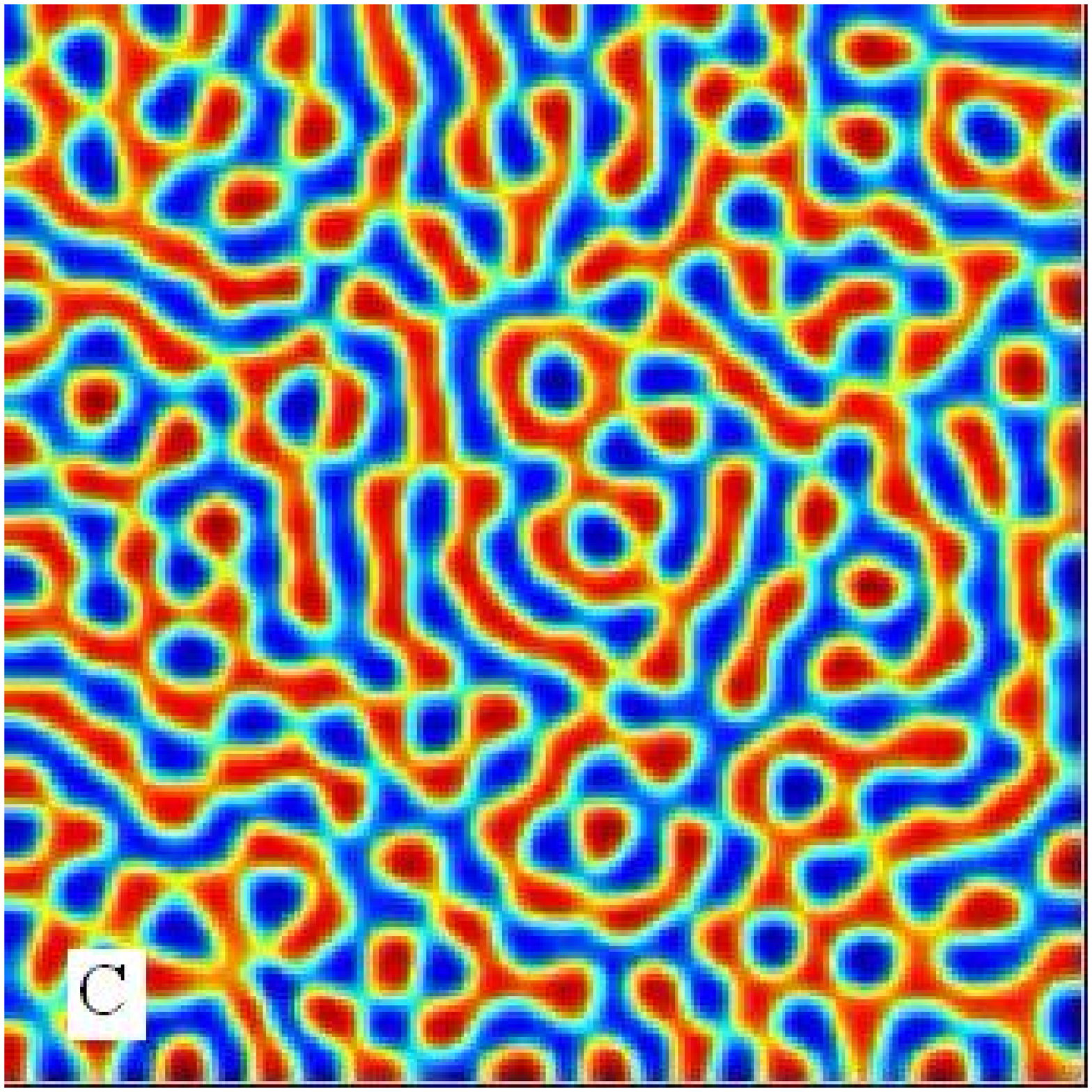}
\caption{\label{figR2}(Color online) Snapshots of contour pictures
of the time evolution of the prey at different instants with
$S=0.9<S_T$. (A) 0 iteration; (B) 5000 iterations; (C) 45000
iterations. [Additional movie format available from the author]}
\end{figure}

We have performed extensive numerical simulations of the spatially
extended model~\eqref{eq:4} in two-dimensional space, and the
qualitative results are shown here. All our numerical simulations
employ the periodic Neumann (zero-flux) boundary conditions with a
system size of $200\times200$ space units and $R=0.5$, $Q=0.6$,
$D_{1}=0.02$, $D_2=0.2$. The spatiotemporal dynamics of a
diffusion-reaction system depends on the choice of initial
conditions, which some authors have considered in connection with
the problem of biological invasion in a few
papers~\cite{JASherratt03281995,NShigesada,Petrovskii:March2001,medvinsky:311},
where the initial conditions are naturally described by finite
functions and the dynamics of the community mainly consists of a
variety of diffusive populational fronts. In general, there are two
initial conditions used for analysis of the spatial extended
systems. One is random spatial distribution of the species, which
seems to be more general from the biological point of view (cf.
Fig.~\ref{figR1}(A) and ~\ref{figR2}(A), ~\ref{figR4}(A)); The other
is a special choice, i.e., taking the species community in a
horizontal layer as decreasing gradually and the vertical
distribution of species homogeneous(cf. Fig.~\ref{figR6}(A)). In
this section we choose the former, and the latter in the next
section (IV. Discussion). The equations~\eqref{eq:4} are solved
numerically in two-dimensional space using a finite difference
approximation for the spatial derivatives and an explicit Euler
method for the time integration with a time stepsize of $\Delta
t=0.01$ and space stepsize $\Delta h=0.25$.

From the analysis of section II and phase-transition bifurcation
diagram (cf. Fig.~\ref{fig1}), the results of computer simulations
show that the type of the system dynamics is determined by the
values of $S$ and $D_1$. We run the simulations until they reach a
stationary state or until they show a behavior that does not seem to
change its characteristics anymore. For different sets of
parameters, the features of the spatial patterns become essentially
different if $S$ exceeds a critical value $S_T$, $S_T$ and $S_W$
respectively, they depend on $D_1$.

Fig.~\ref{figR1} shows the evolution of the spatial pattern of prey
at 0, 5000 and 45000 iterations, with random small perturbation of
the stationary solution $n^*$ and $p^*$ of the spatially homogeneous
systems when $S$ is less than the Turing bifurcation threshold
$S_T$. In this case, one can see that for the system~\eqref{eq:4},
the random initial distribution lead to the formation of a strongly
irregular transient pattern in the domain. After the irregular
pattern form (cf. Fig.~\ref{figR1}(B)) it grows slightly and
``jumps" alternately for a certain time, and finally the chaos
spiral patterns prevail over the whole domain, and the dynamics of
the system does not undergo any further changes (cf.
Fig.~\ref{figR1}(C) and the addition movie for Fig.~\ref{figR1}).

\begin{figure}[htp]
\includegraphics[width=3.0cm]{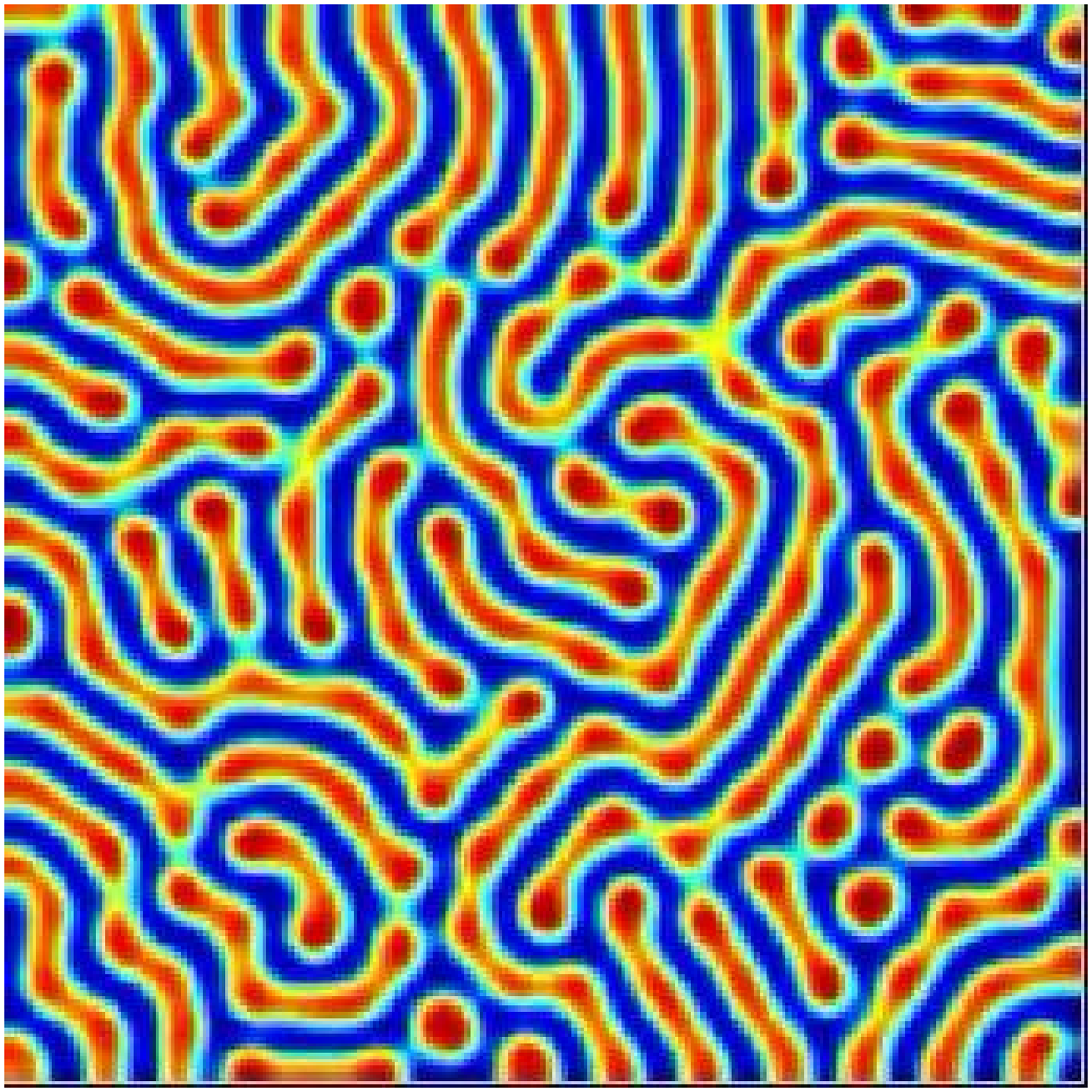}
\caption{\label{figR3}(Color online) Snapshots of contour pictures
of the time evolution of the prey at different instants with
$S_T<S=1.1<S_H$ (45000 iterations). [Additional movie format
available from the author]}
\end{figure}

Figures ~\ref{figR2} and \ref{figR3} show spontaneous formation of
short stripelike and spotted spatial patterns emerge and coexist
stably when the bifurcation parameter $S<S_T$ (Fig.~\ref{figR2}) and
$S_T<S<S_H$ (Fig.~\ref{figR3}). From the snapshots or movies, one
can see that the stripelike spatial patterns arise from the random
initial conditions. After the stripelike patterns form (cf.
Fig.~\ref{figR2}(B)) they grow steadily with time until they reach
certain width---armlength, and the spatial patterns become distinct.
Finally, the stripelike spatial patterns prevail the whole domain
(cf. Fig.~\ref{figR2}(C) and Fig.\ref{figR3}). Comparing the
Fig.~\ref{figR2}(C) with Fig.~\ref{figR3}, we find that the
parameter $S$ is closer to $S_T$, and the stripelike spatial
patterns are more distinct. Here we omit the pre-image of
Fig.~\ref{figR3} as they are similar to the Fig.~\ref{figR2}(A) and
(B). It is easy to see that the stationary patterns are essentially
different from the previous case(cf. Fig.\ref{figR1}).

\begin{figure}[htp]
\includegraphics[width=4cm]{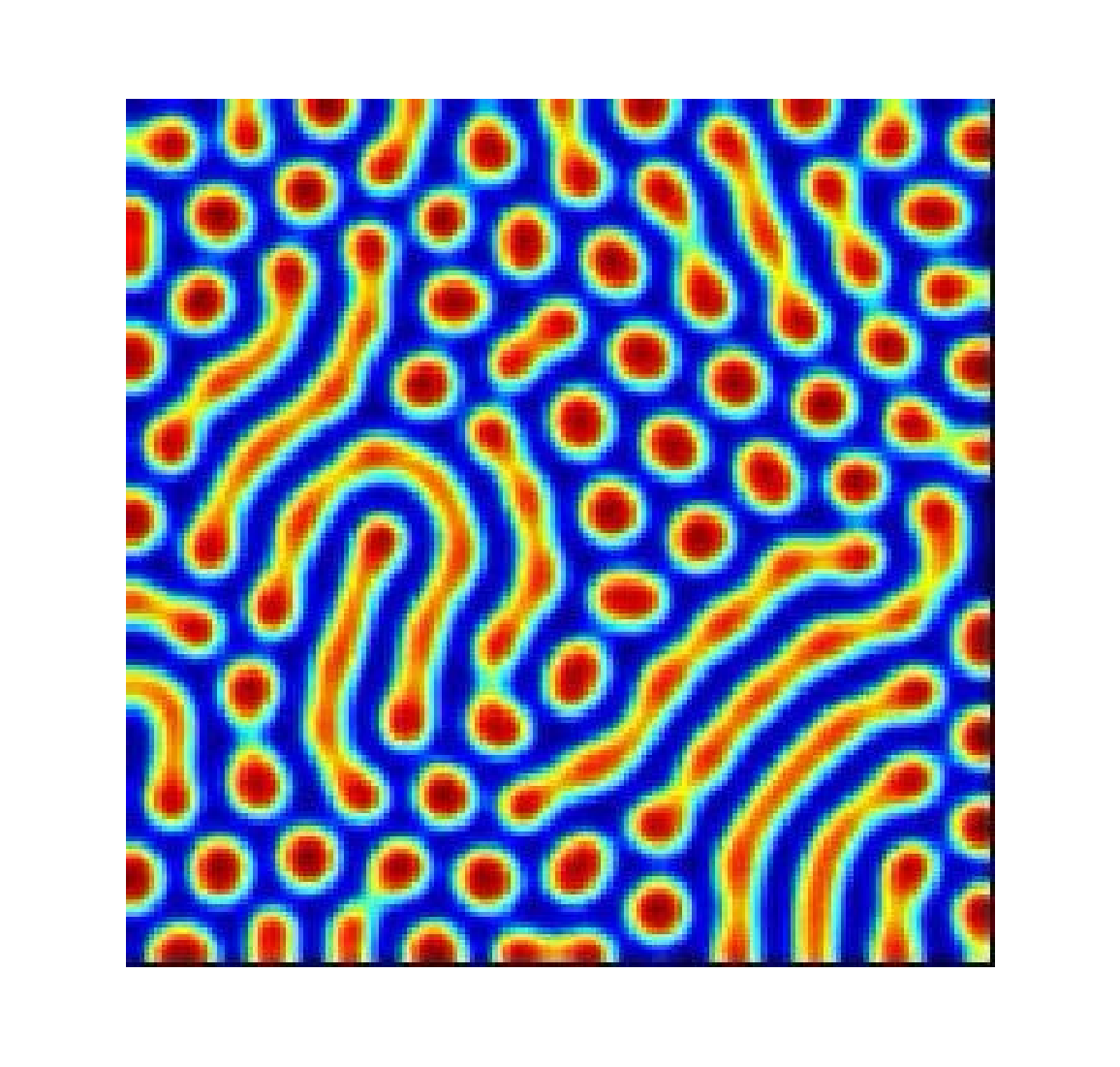}
\caption{\label{figR7}(Color online) Coexistence of stationary
spotted patterns and stripelike patterns of the prey for long time
run with $S_H<S=1.16<S_{W}$. [Additional movie format available from
the author]}
\end{figure}

\begin{figure}[htp]
\includegraphics[width=3cm]{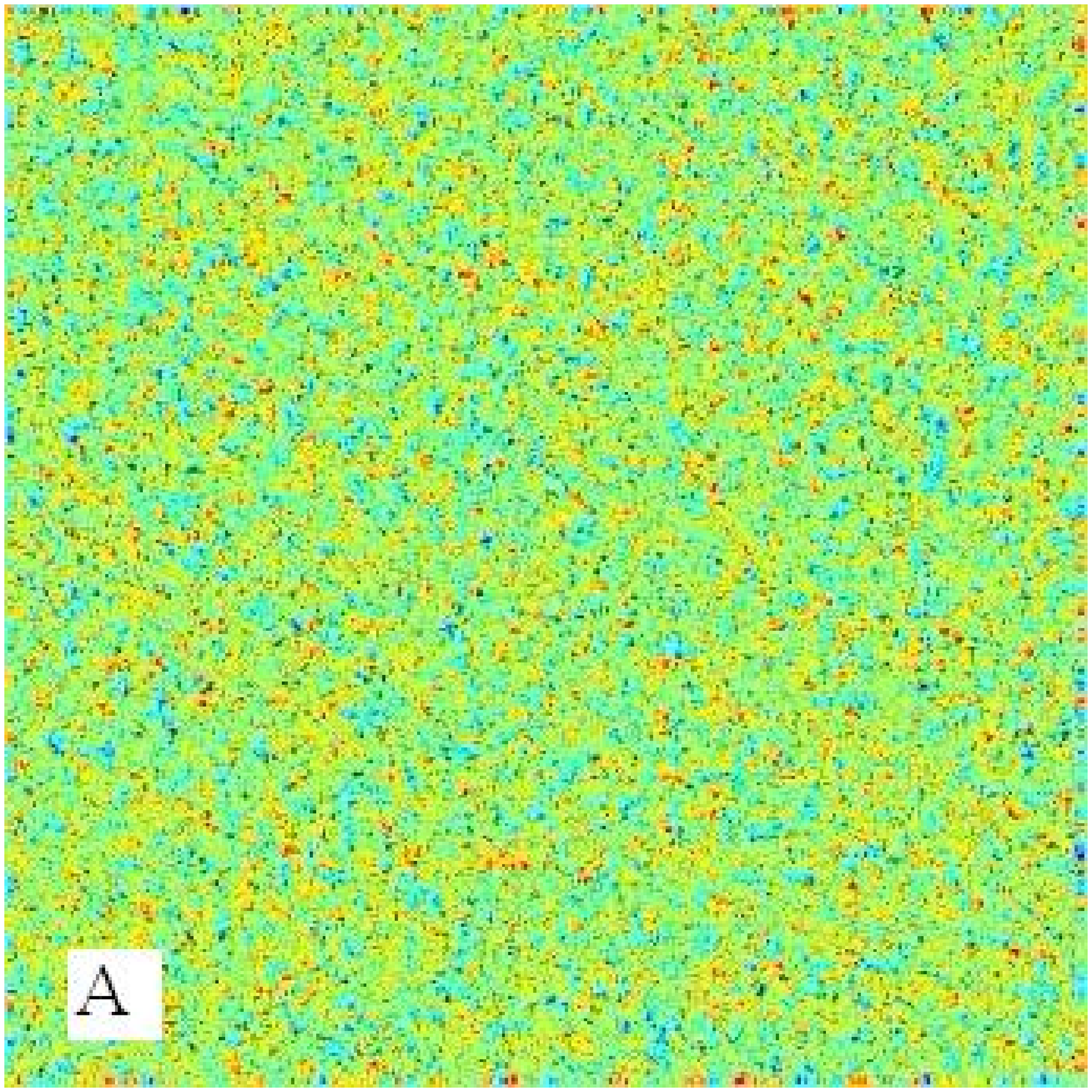}
\includegraphics[width=3cm]{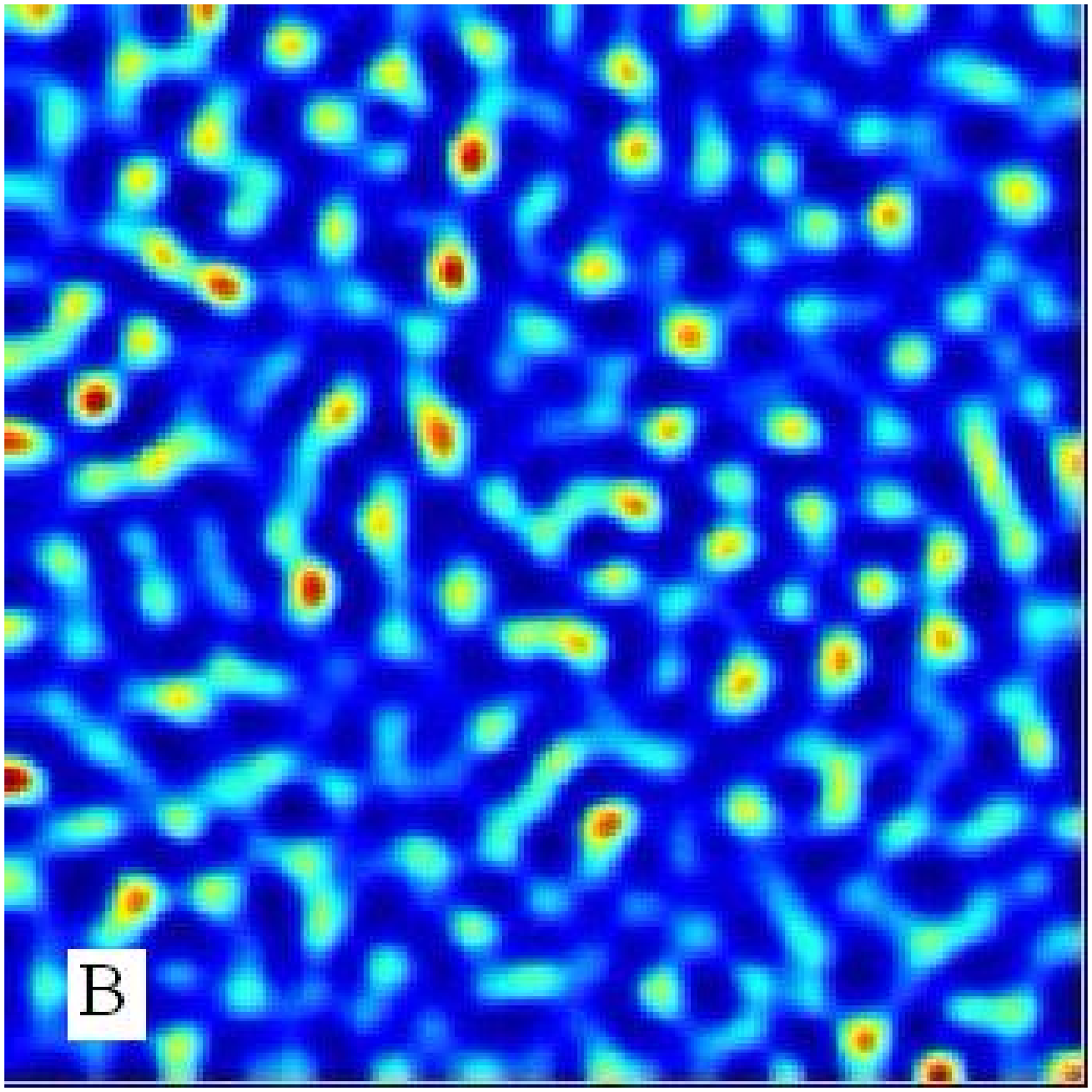}
\includegraphics[width=3cm]{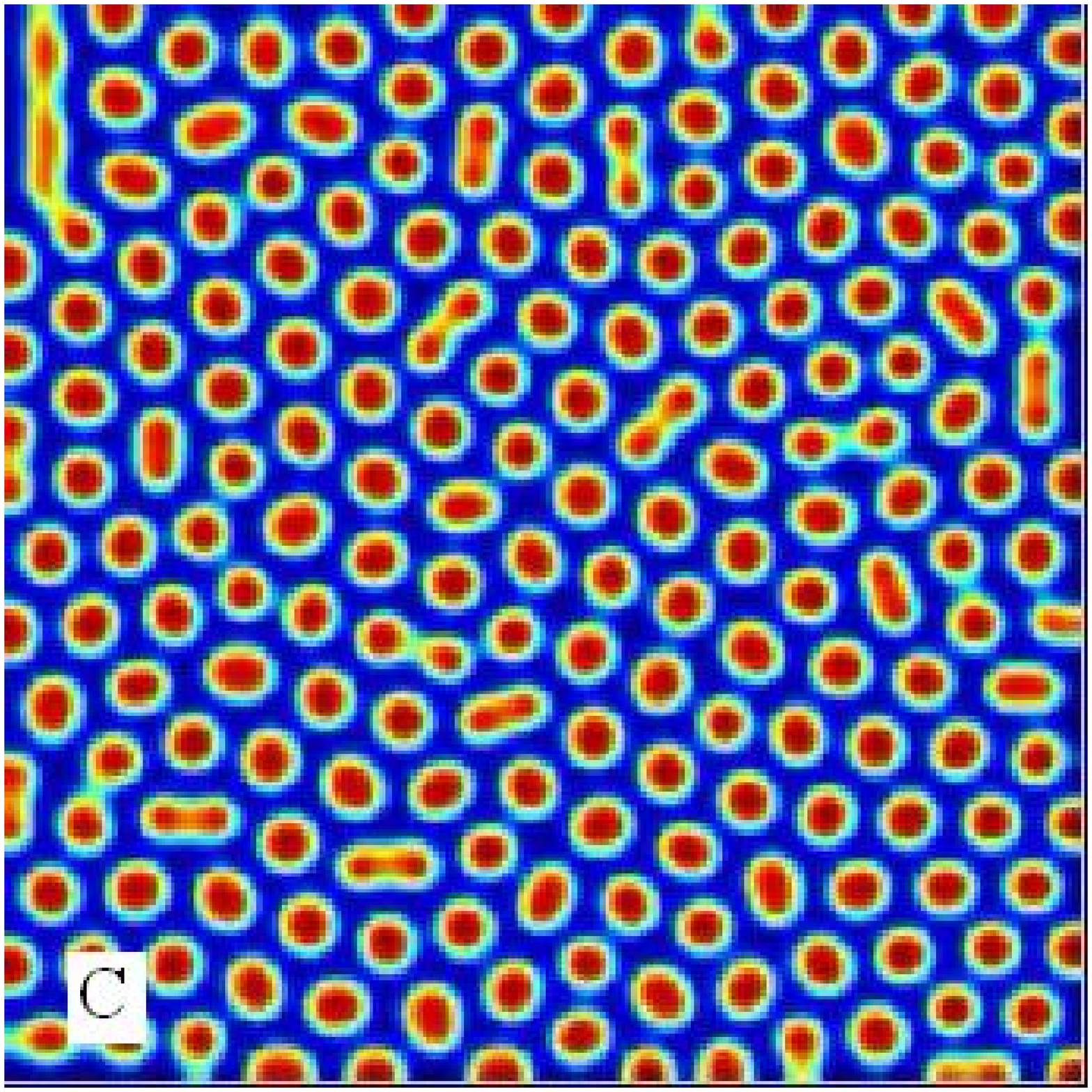}
\caption{\label{figR4}(Color online) Snapshots of contour pictures
of the time evolution of the prey at different instants with
$S=1.2>S_{W}$. (A) 0 iteration; (B) 5000 iterations; (C) 45000
iterations. [Additional movie format available from the author]}
\end{figure}

When $S_H<S=1.16<S_W$, we find that the spotted patterns and the
stripelike patterns coexist in the spatially extended model (cf.
Fig.~\ref{figR7}).

Fig.~\ref{figR4} shows snapshots of prey spatial pattern at 0, 5000
and 45000 iterations for the parameter $S=1.2>S_W$. Although the
dynamics of the system starts from the same initial condition as
previous cases, there is an essential difference for the spatially
extended model (Eq.~\ref{eq:4}). Form Fig.~\ref{figR4}, one can see
that the regular spotted patterns prevails over the whole domain at
last, and the dynamics of the system does not undergo any further
changes (cf. the additional movie for Fig.~\ref{figR4}).

\begin{figure*}[htp]
\includegraphics[width=4cm]{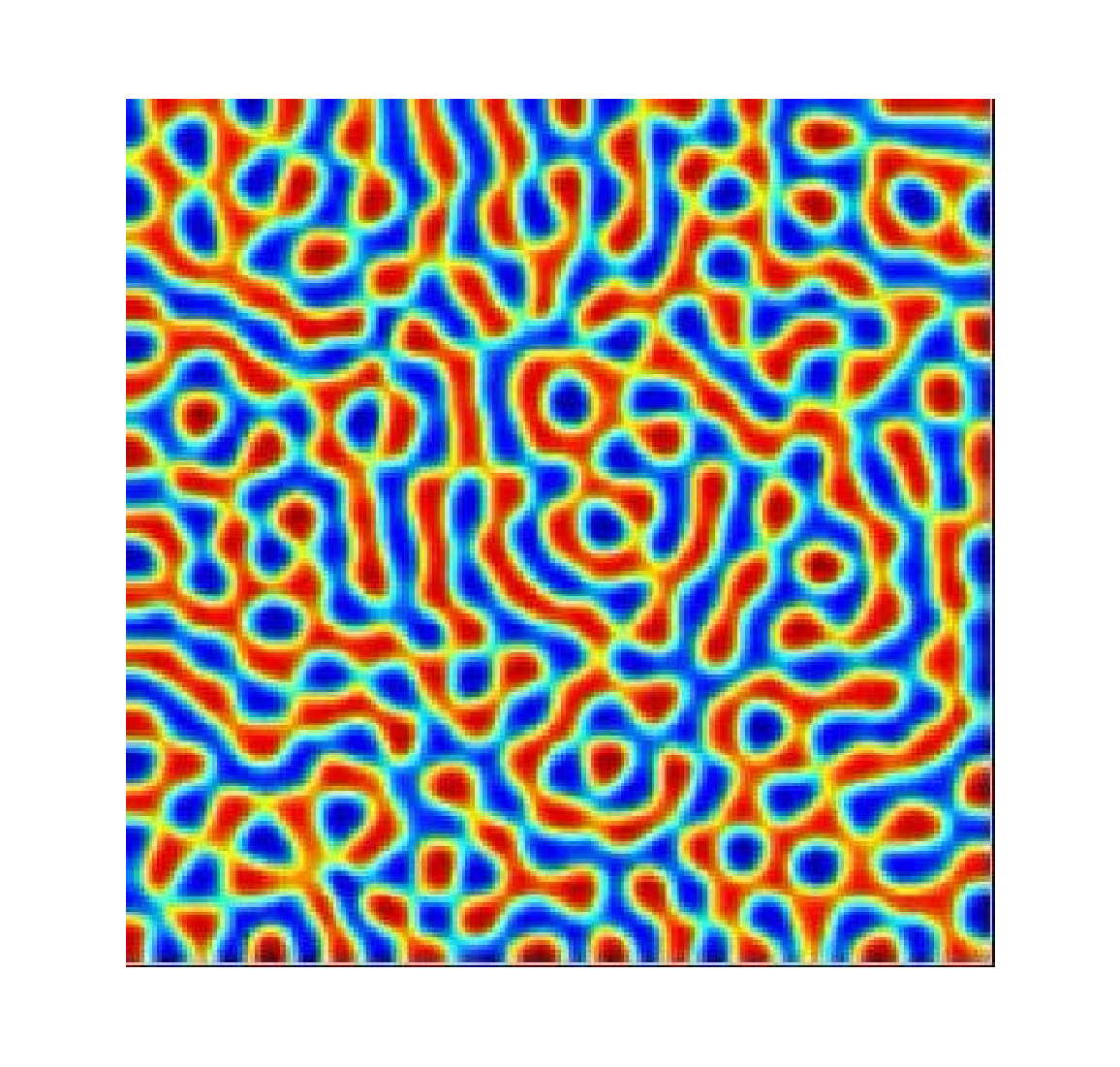}
\includegraphics[width=5.2cm]{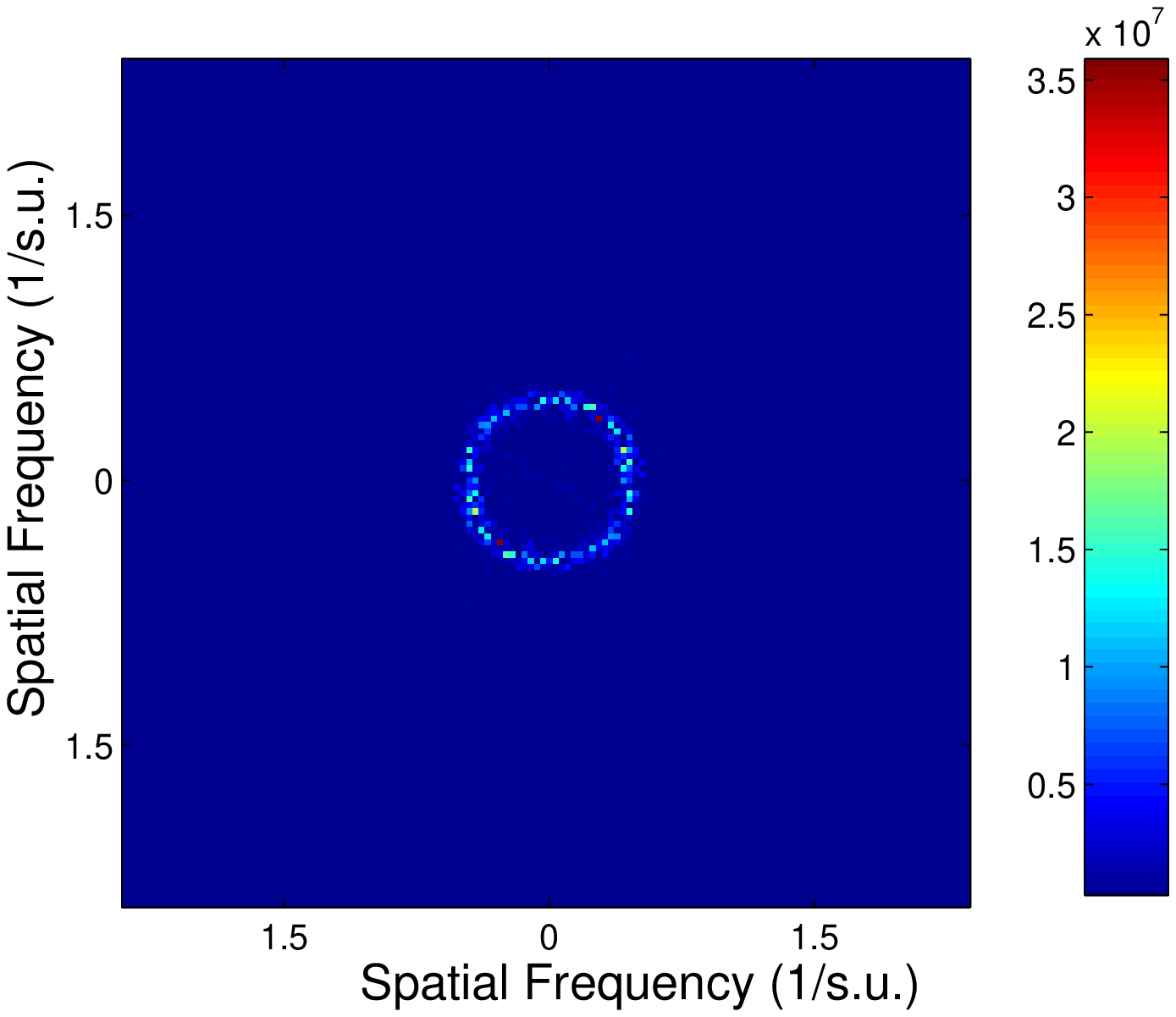}
\includegraphics[width=5.2cm]{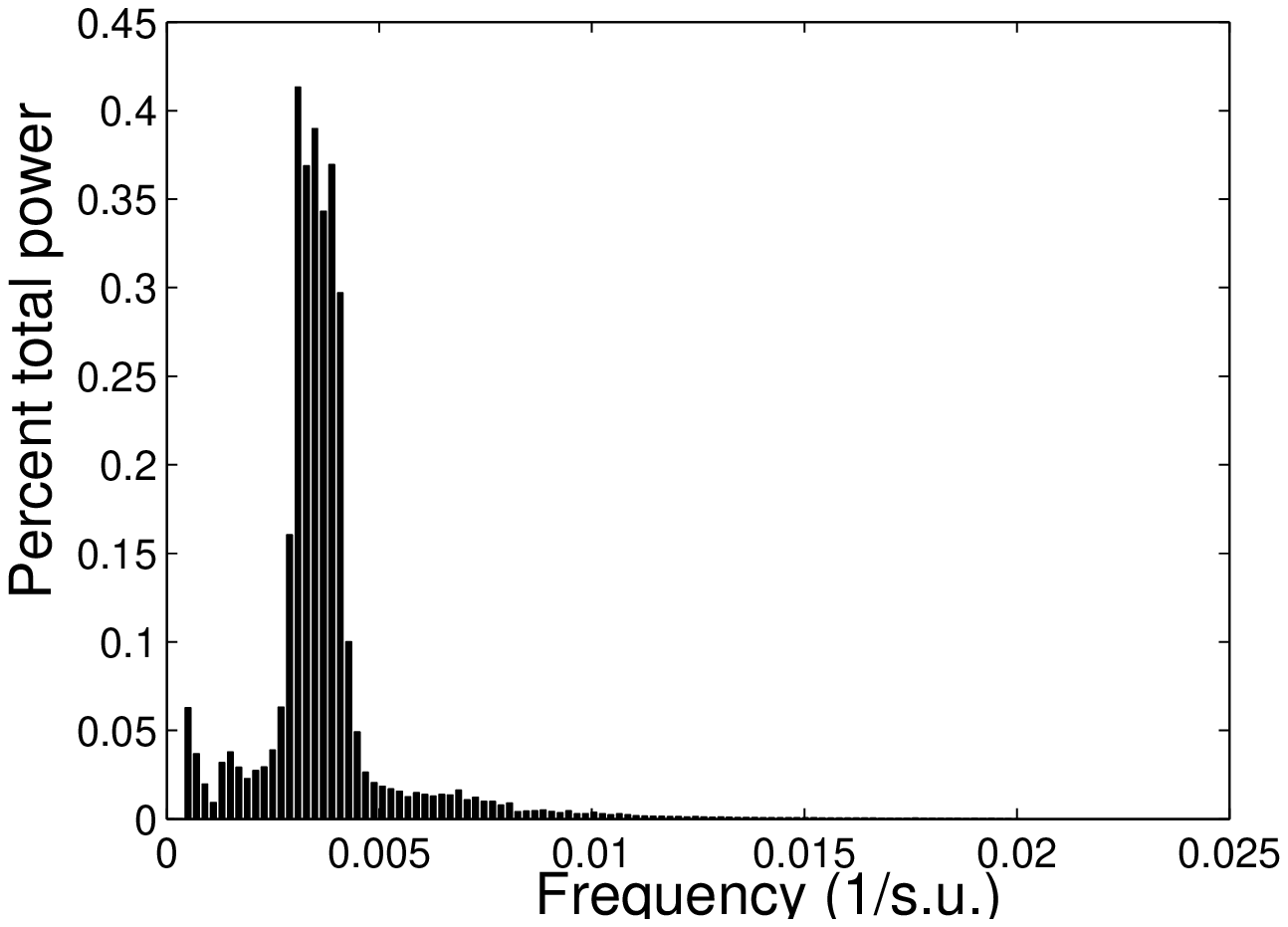}\\
\includegraphics[width=4cm]{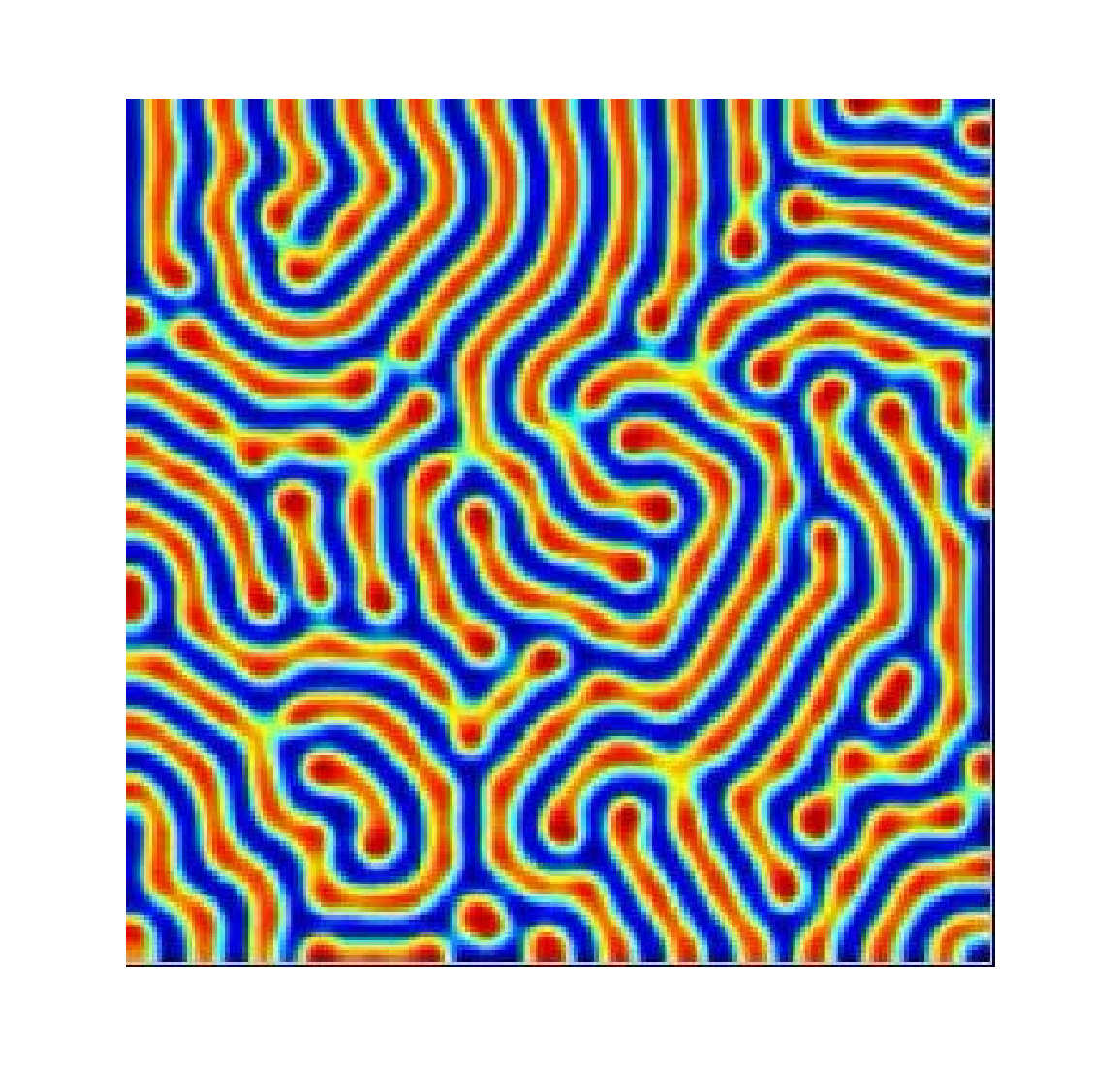}
\includegraphics[width=5.2cm]{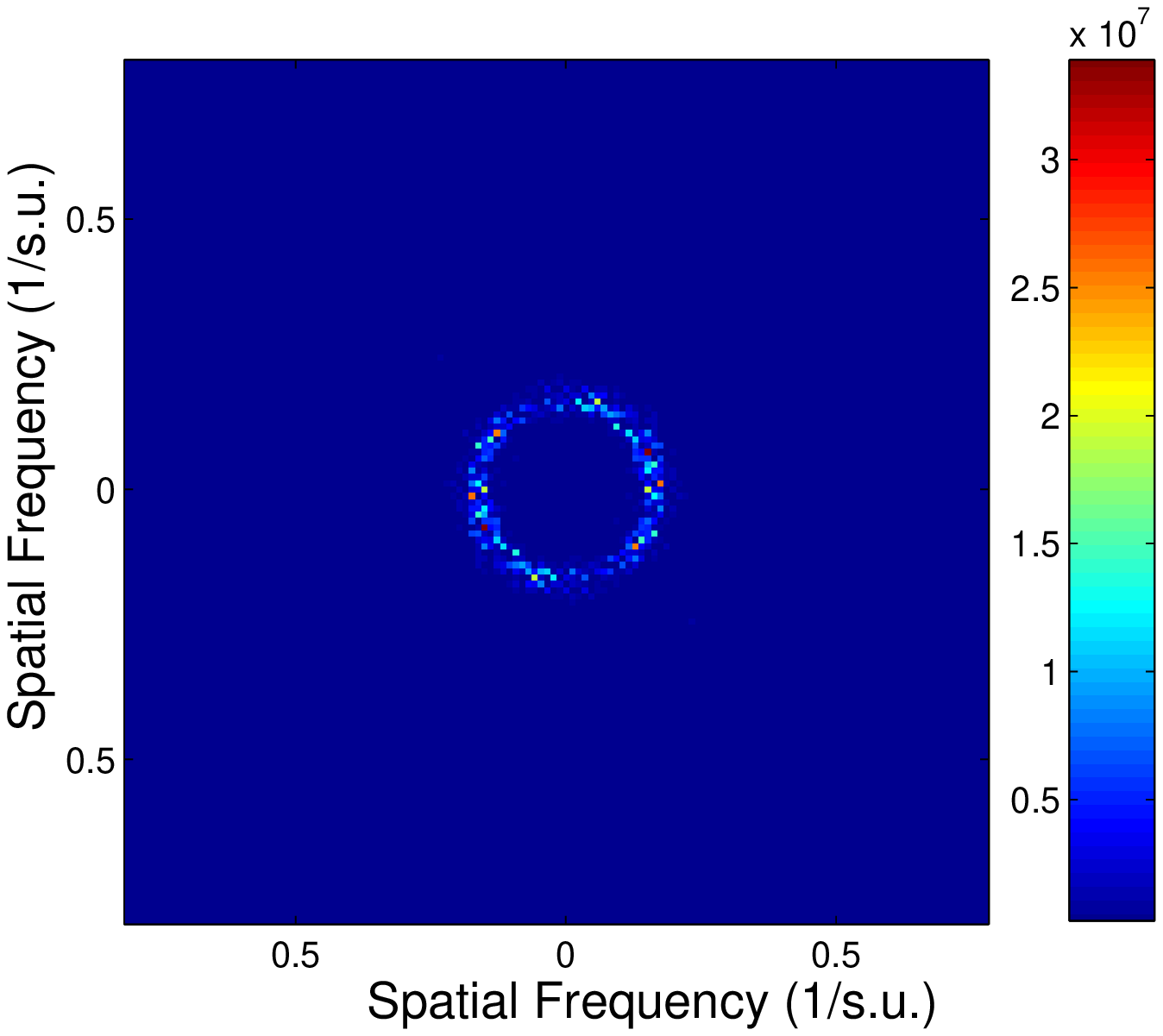}
\includegraphics[width=5.2cm]{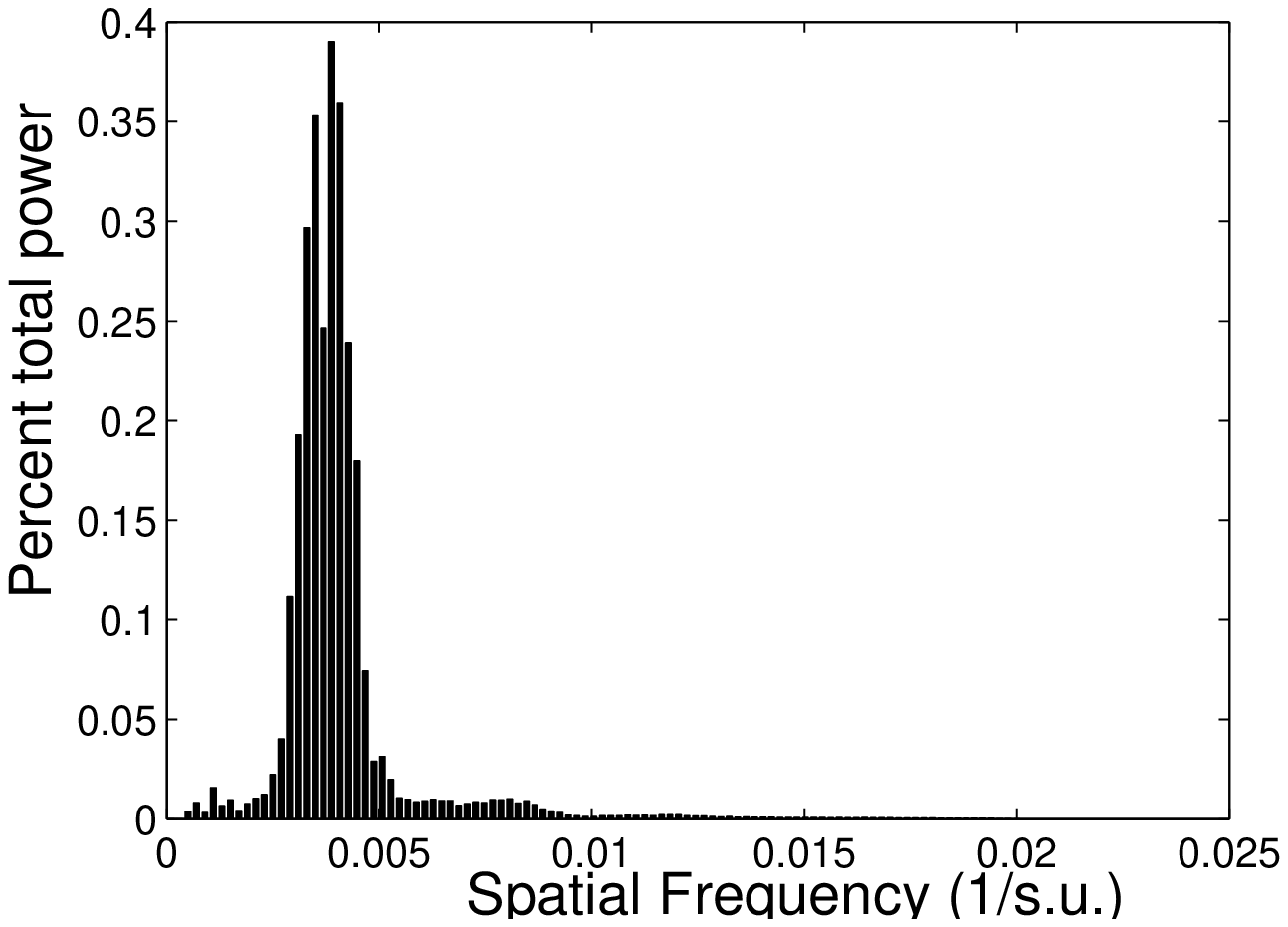}\\
\includegraphics[width=4cm]{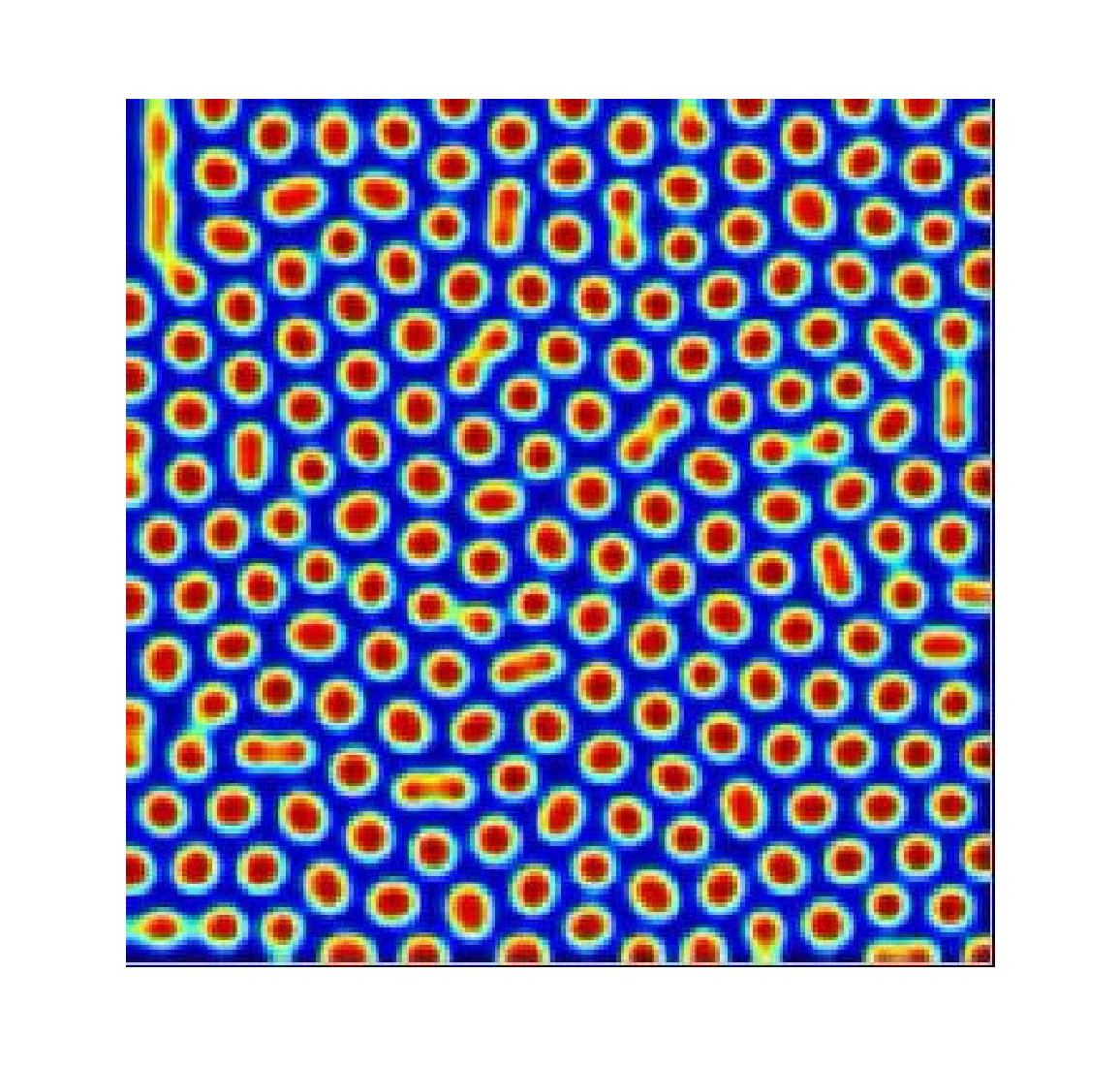}
\includegraphics[width=5.2cm]{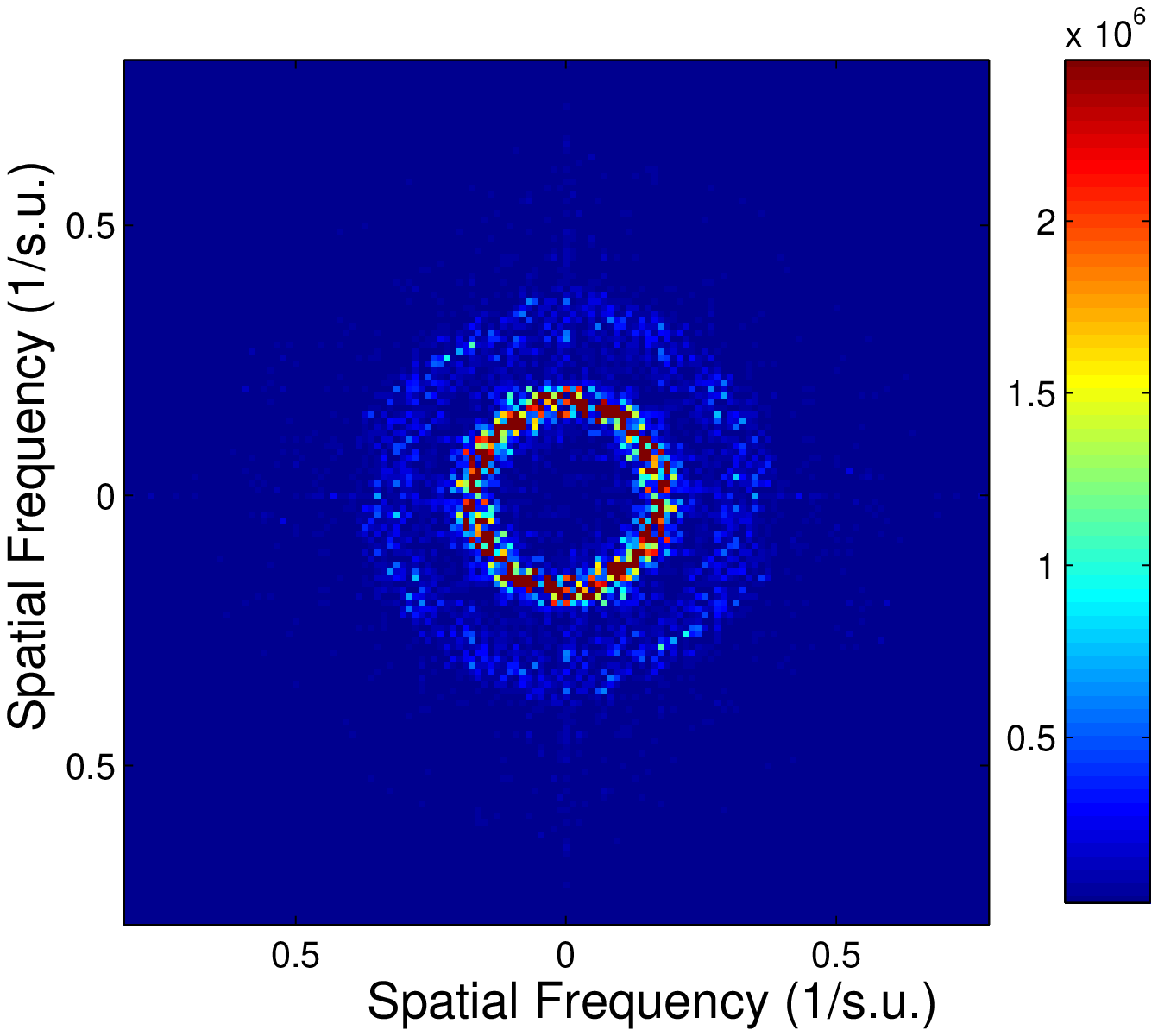}
\includegraphics[width=5.2cm]{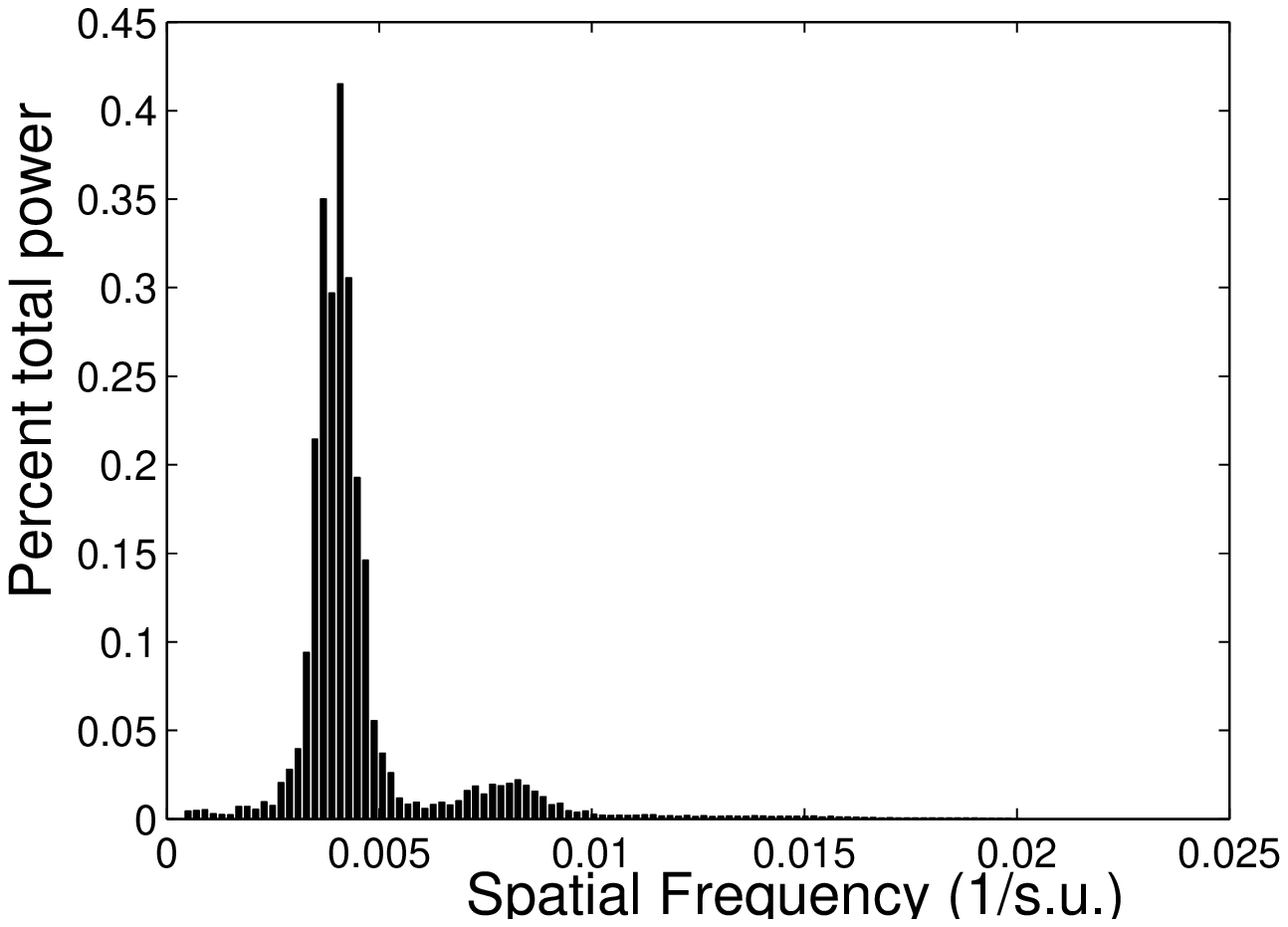}
\caption{\label{figR5}(Color online) Patterns (left-hand column),
 it spatial Fourier transformation (mid-hand column), and radial average of
 the power spectrum (right-hand column). Up-row
$S=0.9$; Mid-row $S=1.1$; Low-row $S=1.2$.}
\end{figure*}

On the other hand, discrete Fourier transform is a basic
mathematical tool used to decompose a signal or image into different
periodic components. It has been widely used for the spatial
patterns~\cite{RichardO.Prum08012003,Prum,RichardO.Prum07152003}. We
have also performed numerical investigations into two-dimensional
space by Fourier spectra. The numerical computation of the Fourier
transform is done by the well-established two-dimensional Fast
Fourier Transform (FFT2) algorithm~\cite{Briggs}. Spatial Fourier
transform of the stripelike and spotted patterns in Figures
\ref{figR2}(c), \ref{figR3} and \ref{figR4}(c) are shown as
Fig.~\ref{figR5}. And digital or digitized transmission electron
micrographs (TEMs) are analyzed by using Matlab (Ver.7.0).

From Fig.~\ref{figR5}, we find that Fig.~\ref{figR2}(C) and
Fig.~\ref{figR3} have the same spatial frequency in the length of
the space unit and presence of one mode with different wave numbers.
On the contrary, Fig.~\ref{figR4}(C) has two modes with different
wave numbers. The spatial frequency and direction of any component
in the power spectrum are given in the length and direction,
respectively, of a vector from the origin to the point on the
circle. The magnitude is depicted by a gray scale or color scale,
but the units are dimensionless values related to the total darkness
of the original images. In Fig.~\ref{figR5}(mid-hand column), short
wavelength, represented by a large circle, corresponds to Turing
structures; longer wavelength, represented by a small white circle,
corresponds to traveling and/or standing waves. This technique can
be particularly appropriate for characterizing quasi-ordered arrays
for Fig.~\ref{figR4}(C).

\section{Discussion and conclusion}

The numerical results correspond perfectly to our theoretical
findings that there are a range of parameters in $S-D_{1}$ plane
where the different spatial patterns emerge (cf. Fig.~\ref{fig1}).
Fig.~\ref{fig1} and the results of simulations present that the
chaos patterns will persist in the spatially extended model
(Eq.~\ref{eq:4}) when the parameters are in the domain I. The
boundary of this domain can be computed numerically and is shown as
the blue line ``Turing" in Fig.~\ref{fig1}. The stationary state of
stripelike patterns exists when the parameters are in the domain II,
where its boundary can also be computed numerically and is shown as
the black line ``Hopf" in Fig.~\ref{fig1}. The periodic spotted
patterns appear in the domain VI, where the boundary can be computed
numerically by the wave bifurcation and is shown as the red line
``Wave" in Fig.~\ref{fig1}. Moreover, there is transverse domain IV
(cf. Fig.~\ref{fig1}) in the system between the stripelike patterns
and spotted patterns, where the spotted patterns and the stripelike
patterns coexist(cf. Fig.~\ref{figR7}).

\begin{figure}[htp]
\includegraphics[width=3cm]{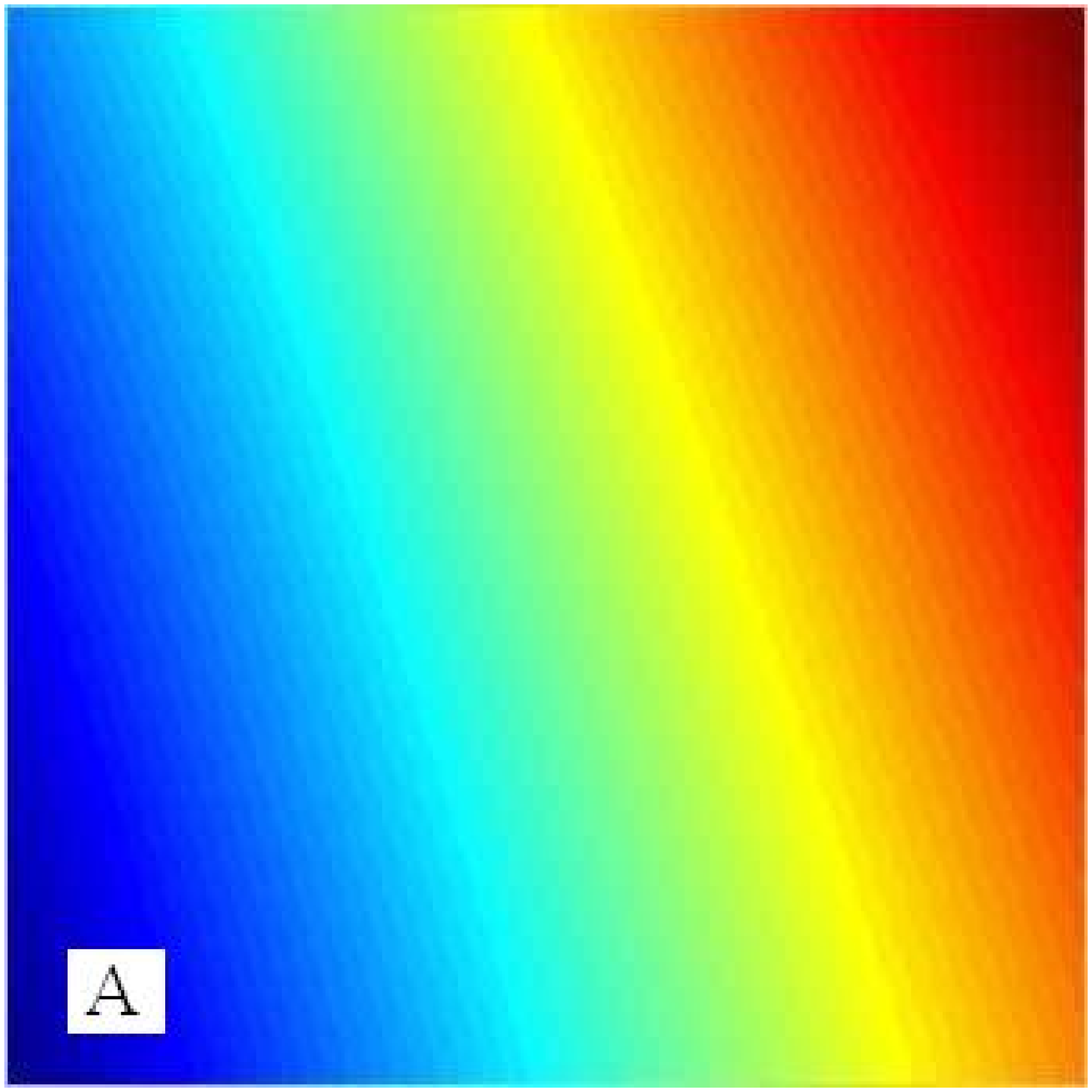}
\includegraphics[width=3cm]{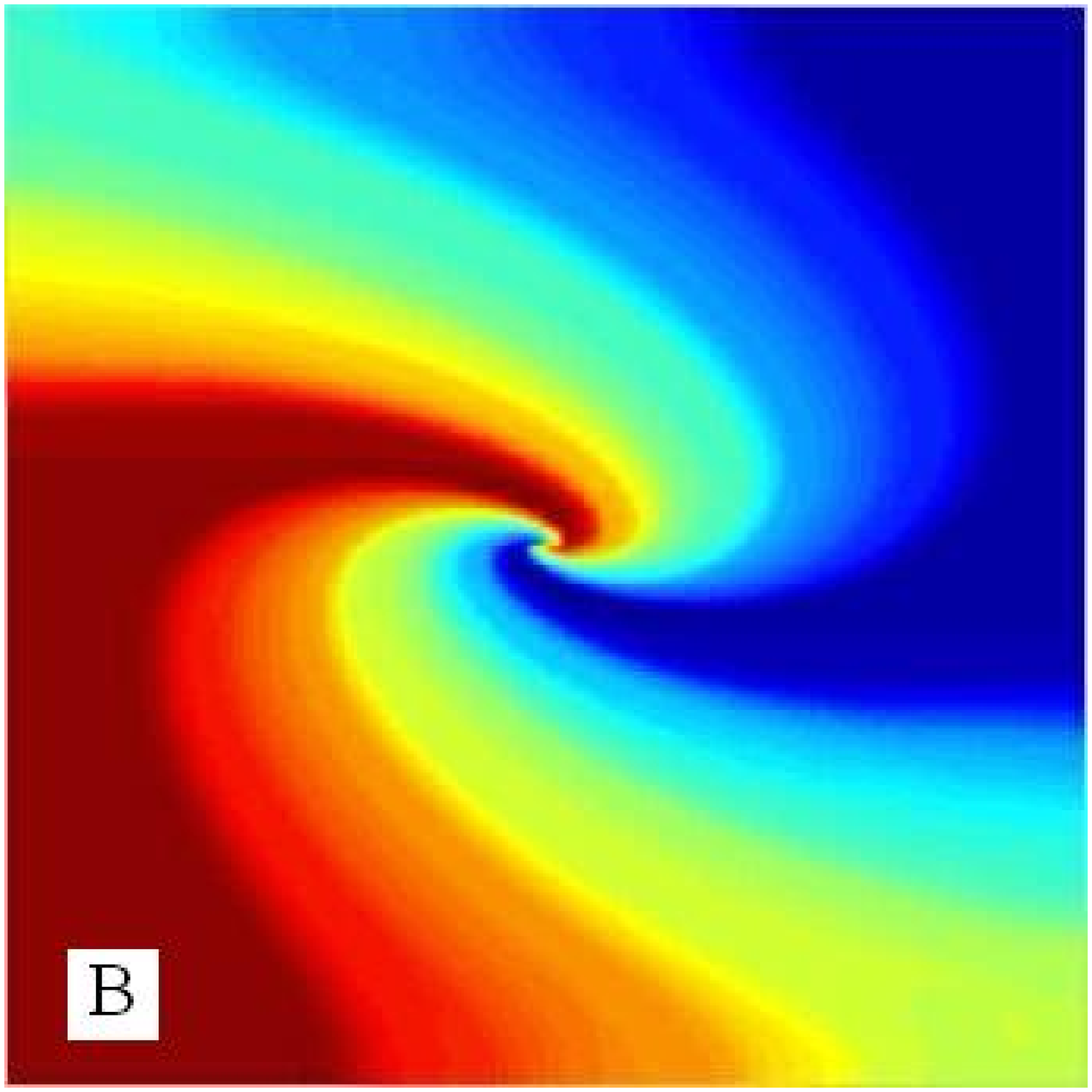}
\includegraphics[width=3cm]{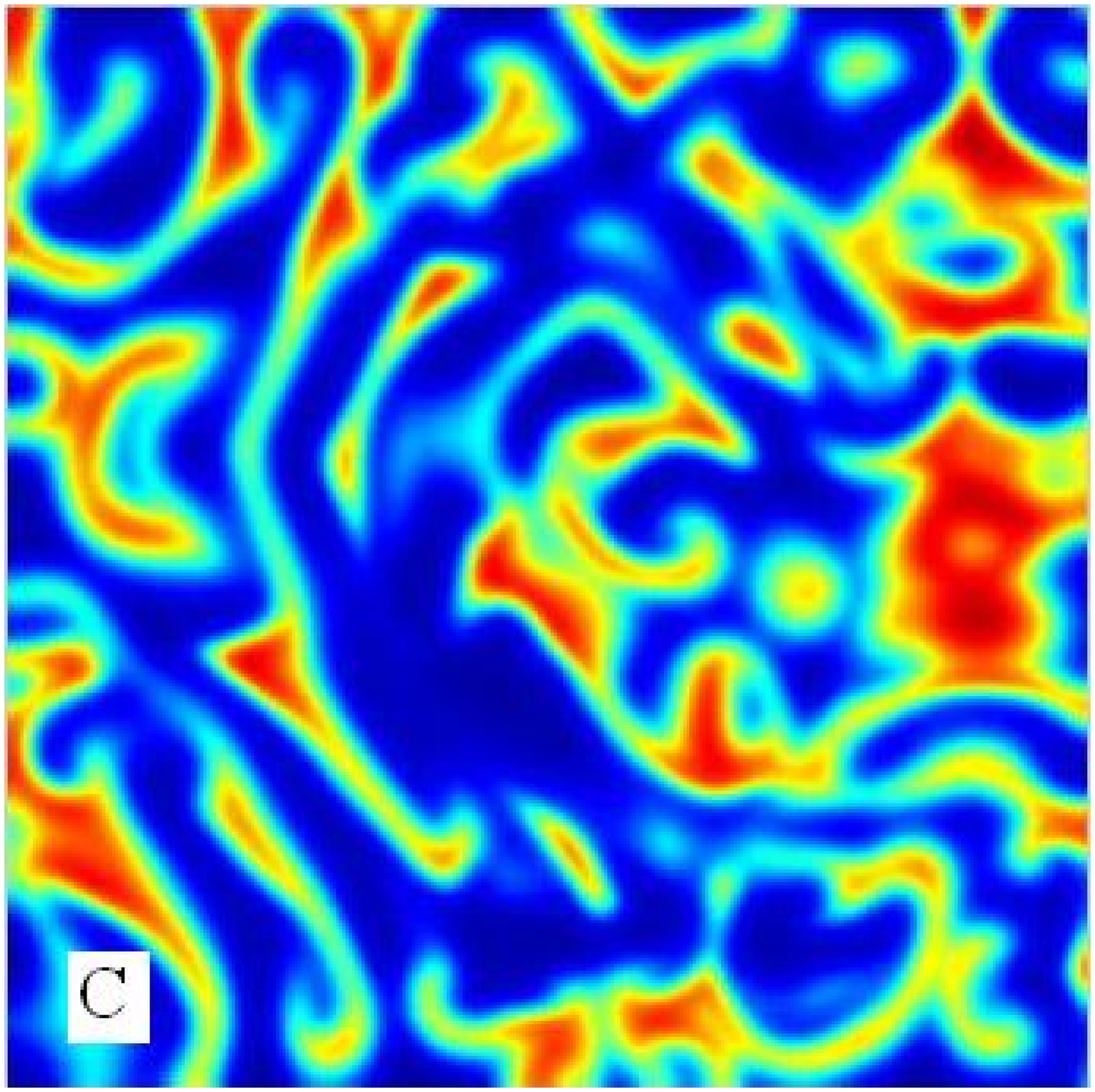}
\caption{\label{figR6}(Color online) Snapshots of contour pictures
of the time evolution of the prey at different instants with the
special initial condition and $S=0.6<S_{T}$. (A) 0 iteration; (B)
1000 iterations; (C) 400000 iterations. [Additional movie format
available from the author]}
\end{figure}

Do the stationary patterns arise dependent on the initial
conditions? We test the different initial conditions for the
spatially extended system, but the final spatial patterns are the
same in qualitative. In those figures we find that the spatial chaos
patterns come from the destruction of the spirals, when we choose
the special initial condition (cf. Fig.~\ref{figR6}) in the domain
I. This phenomenon coheres with the results of the study in
Refs.~\cite{medvinsky:311,Gurney}.

We have presented a theoretical analysis of evolutionary processes
that involves organisms distribution and their interaction of
spatially distributed population with local diffusion. Our analysis
and numerical simulations reveal that the typical dynamics of
population density variation is the formation of isolated groups
(stripelike or spotted or coexistence of both). This process depends
on several parameters, including $S$, $D_1$ and $D_2$. The field
meaning of our results may be found in the dynamics of an aquatic
community which is affected by the existence of relatively stable
mesoscale inhomogeneity in the field of ecologically significant
factors such as water temperature, salinity and biogen
concentration.

In Ref.~\cite{medvinsky:311}, the authors explained the field
meaning by using aquatic community in the ocean (cf.
Fig.~\ref{figR6}). Our study shows that the spatially extended model
(Eq.~\ref{eq:4}) has not only more complex dynamic patterns in the
space, but also chaos patterns and spiral waves, so it may help us
better understand the dynamics of an aquatic community in a real
marine environment. It is also important to distinguish between
``intrinsic" patterns, i.e., patterns arising due to trophic
interactions like those considered above, and ``forced" patterns
induced by the inhomogeneity of the environment. The physical nature
of the environmental heterogeneity, and thus the value of the
dispersion of varying quantities and typical times and lengths, can
be essentially different in different cases. Neuhauser and
Pacala~\cite{Neuhauser01171997} formulated the Lotka-Volterra model
as a spatial model. They found the striking result that the
coexistence of patterns is actually harder to get in the spatial
model than in the non-spatial one. One reason can be traced to how
local interactions between individual members of the species are
represented in the model. In this thesis, our results show that the
ratio-dependent predator-prey model (Eq.~\ref{eq:4}) also represents
rich spatial dynamics, such as chaos spiral patterns, stripelike
patterns, spotted patterns, coexistence of both stripelike and
spotted patterns, etc. It will be useful for studying the dynamic
complexity of ecosystems.

\begin{acknowledgments}
This work was supported by the National Natural Science Foundation
of China under Grant No. 10471040 and the Natural Science Foundation
of Shan'xi Province Grant No. 2006011009.
\end{acknowledgments}

\bibliographystyle{apsrev}

\end{document}